\documentclass[aps,prd,11pt,twoside,tightenlines,nofootinbib,showpacs,preprint,superscriptaddress]{revtex4-1}
\usepackage[colorlinks=true]{hyperref}
\usepackage{xcolor}
\hypersetup{colorlinks=true, citecolor=blue, urlcolor=blue, linkcolor=blue}
\usepackage{amsmath,amssymb}
\usepackage[final]{graphicx}
\usepackage{subcaption}
\usepackage{slashed}
\usepackage{float}
\usepackage[toc,page]{appendix}

\begin{document}
	
\arraycolsep1.5pt
\newcommand{\Ima}{\textrm{Im}}
\newcommand{\Rea}{\textrm{Re}}
\newcommand{\mev}{\textrm{ MeV}}
\newcommand{\gev}{\textrm{ GeV}}

\title{Are the $a_{0}(1710)$ or $a_{0}(1817)$ resonances in the $D_{s}^{+} \rightarrow K_{S}^{0}K^{+}\pi^{0}$ decay?}

\author{Zhong-Yu Wang}
\affiliation{School of Physical Science and Technology, Lanzhou University, Lanzhou 730000, China}
\affiliation{Lanzhou Center for Theoretical Physics, Key Laboratory of Theoretical Physics of Gansu Province, and Key Laboratory of Quantum Theory and Applications of MoE, Lanzhou University, Lanzhou, Gansu 730000, China}

\author{Yu-Wen Peng}
\affiliation{School of Physics and Electronics, Hunan Key Laboratory of Nanophotonics and Devices, Central South University, Changsha 410083, China}

\author{Jing-Yu Yi}
\affiliation{School of Physics and Electronics, Hunan University, Changsha 410082, China}

\author{W. C. Luo}
\email{luo.wenchen@csu.edu.cn}
\affiliation{School of Physics and Electronics, Hunan Key Laboratory of Nanophotonics and Devices, Central South University, Changsha 410083, China}

\author{C. W. Xiao}
\email{xiaochw@gxnu.edu.cn}
\affiliation{School of Physics and Electronics, Hunan Key Laboratory of Nanophotonics and Devices, Central South University, Changsha 410083, China}
\affiliation{Department of Physics, Guangxi Normal University, Guilin 541004, China}

\date{\today}

\begin{abstract}

The BESIII Collaboration claimed that a new $a_{0}(1817)$ resonance was found in the recent results of the $D_{s}^{+} \rightarrow K_{S}^{0}K^{+}\pi^{0}$ decay. 
For this decay process, we perform a unitary amplitude to analyze the contributions of the states $a_{0}(980)^{+}$ and $a_{0}(1710)^{+}$ with the final state interactions. 
Considering the Cabibbo-favored external and internal $W$-emission mechanisms at the quark level, and the contributions of the resonances $a_{0}(980)^{+}$, $a_{0}(1710)^{+}$ in the $S$-wave and $\bar{K}^{*}(892)^{0}$, ${K}^{*}(892)^{+}$ in the $P$-wave, the recent experimental data of the $K_{S}^{0}K^{+}$ invariant mass spectrum from the BESIII Collaboration can be described well.
In our results, the states $a_{0}(980)$ and $a_{0}(1710)$ are dynamically generated from the final state interactions of $K\bar{K}$ and $K^{*}\bar{K}^{*}$, respectively, which support the molecular nature for them.
Moreover, some obtained branching fractions are in agreement with the experimental measurements.

\end{abstract}
\pacs{}

\maketitle

\section{Introduction}

Recently, the BESIII Collaboration performed the amplitude analysis of the decay $D_{s}^{+} \rightarrow K_{S}^{0}K^{+}\pi^{0}$, and  reported the branching fraction $\mathcal{B}\left(D_s^{+} \rightarrow K_S^0 K^{+} \pi^0\right)=\left(1.46 \pm 0.06 \pm 0.05\right) \%$ \cite{BESIII:2022npc}, which was consistent with the measurement of the CLEO Collaboration \cite{CLEO:2013bae}. The isovector partner of the $f_{0}(1710)$, a state $a_{0}(1710)^{+}$ was observed in the $K_{S}^{0}K^{+}$ invariant mass spectrum of the decay $D_{s}^{+} \rightarrow K_{S}^{0}K^{+}\pi^{0}$ \cite{BESIII:2022npc} 
\footnote{In the published version, they assumed to be a new $a_{0}(1817)$ resonance.}, 
of which the mass and width were measured as,
\begin{equation*}
	M_{a_{0}(1710)} = (1.817\pm0.008\pm0.020)\gev,~
	\Gamma_{a_{0}(1710)}= (0.097\pm0.022\pm0.015)\gev. 
\end{equation*}
In fact, previously, the BABAR Collaboration performed the Dalitz plot analyses of $\eta_{c}\rightarrow\eta\pi^{+}\pi^{-}$ decay and found a new state $a_{0}(1700)$ in the $\pi\eta$ invariant mass spectrum \cite{BaBar:2021fkz}, 
\begin{equation*}
	M_{a_{0}(1700)} = (1.704\pm0.005\pm0.002)\gev, ~
	\Gamma_{a_{0}(1700)}= (0.110\pm0.015\pm0.011)\gev, 
\end{equation*}
which might also be the same state as $a_{0}(1710)$ and corroborated the evidence found in Ref. \cite{BaBar:2018uqa}.
In Ref. \cite{BESIII:2021anf}, a peak around $1.710$ GeV was observed in the $K_{S}^{0}K_{S}^{0}$ mass distribution in the decay $D_{s}^{+} \rightarrow \pi^{+}K_{S}^{0}K_{S}^{0}$ by the BESIII Collaboration. Due to the strong overlap and common quantum numbers $J^{PC}=0^{++}$, the states $a_{0}(1710)$ and $f_{0}(1710)$ were not distinguished, and then together denoted as $S(1710)$, where the mass and width were determined as \cite{BESIII:2021anf},
\begin{equation*}
	M_{S(1710)} = (1.723\pm0.011\pm0.002)\gev, ~
	\Gamma_{S(1710)}= (0.140\pm0.014\pm0.004)\gev. 
\end{equation*}
From these reported results of the BESIII \cite{BESIII:2022npc,BESIII:2021anf} and BABAR \cite{BaBar:2021fkz} Collaborations, the extracted Breit-Wigner masses of $a_{0}(1710)$ are quite different.
Actually, these experimental results have extraordinary significance, because searching for the $a_{0}(1710)$ is crucial to understand the nature of its isoscalar partner state $f_{0}(1710)$. 
In the present work, based on the recent results of the BESIII Collaboration \cite{BESIII:2022npc}, we try to understand the properties of the state $a_{0}(1710)$ by exploiting the final state interaction formalism.

In the quark model, the $f_{0}(1710)$ was interpreted as an $I^{G}(J^{PC})=0^{+}(0^{++})$ light scalar meson by the Godfrey and Isgur model \cite{Godfrey:1985xj}, which should also have an isovector partner at $1.78$ GeV. Similar results were obtained in Ref. \cite{Segovia:2008zza} with a constituent quark model. 
However, the $f_{0}(1710)$ mainly decays to the channels $K\bar{K}$ and $\eta\eta$, indicating that it may have large $s\bar{s}$ quarks components \cite{Chanowitz:2005du,Chao:2007sk}.
The $f_{0}(1710)$ was also regarded as a scalar glueball or containing a large glueball components in Refs. \cite{Close:2005vf,Giacosa:2005zt,Cheng:2006hu,Albaladejo:2008qa,Gui:2012gx,Janowski:2014ppa,Ochs:2013gi,Cheng:2015iaa,Klempt:2021wpg}, which were supported by the experimental resuts of the BESIII Collaboration \cite{BESIII:2022riz,BESIII:2022iwi}. Searching for the isovector partner of the $f_{0}(1710)$ is the key to identify whether it is a scalar glueball.
On the other hand, based on the chiral unitary approach (ChUA) \cite{Oller:1997ti,Oset:1997it,Oller:2000ma,Oller:2000fj,Oset:2008qh}, the $f_{0}(1710)$ was dynamically generated in the interactions of vector mesons and assumed to be a molecular state of $K^{*}\bar{K}^{*}$ in Ref. \cite{Geng:2008gx}, where its pole located at $1.726$ GeV and another $a_0$ state at $1.78$ GeV with isospin $I=1$ was predicted. 
Similar results were obtained in the extended research works of \cite{Du:2018gyn,Wang:2021jub}. 
Indeed, this $a_0$ state at $1.78$ GeV was arranged as the new found $a_{0}(1710)$ state in a further work of \cite{Wang:2022pin}, which was also a bound state of $K^{*}\bar{K}^{*}$, a molecular state. 
More discussions about the molecular states can be referred to the review of Ref. \cite{Guo:2017jvc}.

Furthermore, based on the results from the BESIII Collaboration \cite{BESIII:2021anf,BESIII:2020ctr}, Ref. \cite{Dai:2021owu} studied the decay modes $D_{s}^{+} \rightarrow \pi^{+}K^{+}K^{-}$, $\pi^{+}K^{0}\bar{K}^{0}$, and $\pi^{0}K^{+}\bar{K}^{0}$, where the $f_{0}(1710)$ and $a_{0}(1710)$ states were dynamically generated from the final state interactions of $K^{*}\bar{K}^{*}$, and the branching ratio of $D_{s}^{+} \rightarrow \pi^{0}K^{+}K_{S}^{0}$ reaction was predicted. 
Analogously, the decay $D_{s}^{+} \rightarrow \pi^{+}K_{S}^{0}K_{S}^{0}$ was investigated in details in Ref. \cite{Zhu:2022wzk}, 
where the $K_{S}^{0}K_{S}^{0}$ and $\pi^{+}K_{S}^{0}$ invariant mass distributions were calculated with the resonance contributions of the scalar $f_{0}(1710)$ and the isovector partner $a_{0}(1710)$, and the results obtained were consistent with the measurements from the BESIII Collaboration \cite{BESIII:2021anf}. 
However, in Ref. \cite{Guo:2022xqu}, the $a_{0}(1710)$ state, newly observed by the BESIII Collaboration  \cite{BESIII:2022npc}, was renamed as $a_{0}(1817)$, which was regarded as the isovector partner of the $X(1812)$ found in Ref. \cite{BES:2006vdb} and classified into the isovector scalar meson family according to the standard Regge trajectory.

Therefore, it is meaningful to understand the nature of the state $a_{0}(1710)$, which is critical for further revealing the property of its isoscalar partner state $f_{0}(1710)$.
The latest experimental measurement of the decay $D_{s}^{+} \rightarrow K_{S}^{0}K^{+}\pi^{0}$ by the BESIII Collaboration \cite{BESIII:2022npc} gives us an opportunity to identify the nature of the $a_{0}(1710)$. 
In the present work, with the framework of the ChUA, we investigate the resonance contributions of the process $D_{s}^{+} \rightarrow K_{S}^{0}K^{+}\pi^{0}$ based on the final state interactions, where the states $a_{0}(980)^{+}$ and $a_{0}(1710)^{+}$ are dynamically generated in the coupled channel interactions of the channels $K\bar{K}$ and $K^{*}\bar{K}^{*}$.
In the interactions of coupled channels, both the pseudoscalar and vector channels are considered, where five channels $K^{*}\bar{K}^{*}$, $\rho\omega$, $\rho\phi$, $K\bar{K}$, and $\pi\eta$, are involved. 
To describe the invariant mass spectra, we also take into account the contributions from the $\bar{K}^{*}(892)^{0}$ and ${K}^{*}(892)^{+}$ in the $P$-wave, which play a crucial role in the intermediate processes $D_{s}^{+}\rightarrow \bar{K}^{*}(892)^{0}K^{+}$ and ${K}^{*}(892)^{+}K_{S}^{0}$, but omit the contribution of the resonance $\bar{K}^{*}(1410)^{0}$, of which the contribution was small as implied in Ref. \cite{BESIII:2022npc}.
The manuscript is organized as follows. In Sec. \ref{sec:Formalism}, we present the theoretical formalism of the decay $D_{s}^{+} \rightarrow K_{S}^{0}K^{+}\pi^{0}$ with the final state interaction. Next, our results are shown in Sec. \ref{sec:Results}. A short conclusion is made in Sec. \ref{sec:Conclusions}.

\section{Formalism}
\label{sec:Formalism}

\begin{figure}[htbp]
	\begin{subfigure}{0.45\textwidth}
		\centering
		\includegraphics[width=1\linewidth,trim=150 530 180 120,clip]{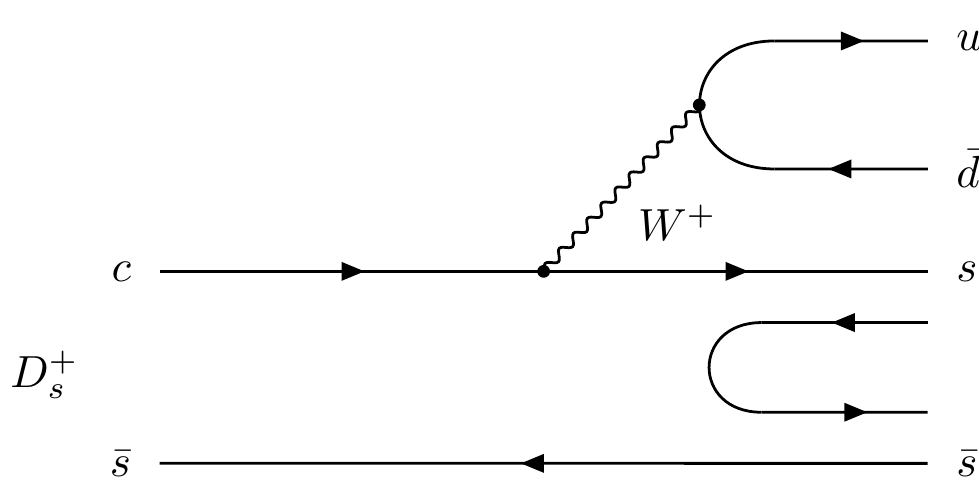} 
		\caption{\footnotesize $s\bar{s}$ quark pair hadronizes into a final meson pair.}
		\label{fig:Feynman1}
	\end{subfigure}
	\quad
	\quad
	\begin{subfigure}{0.45\textwidth}  
		\centering 
		\includegraphics[width=1\linewidth,trim=150 530 180 120,clip]{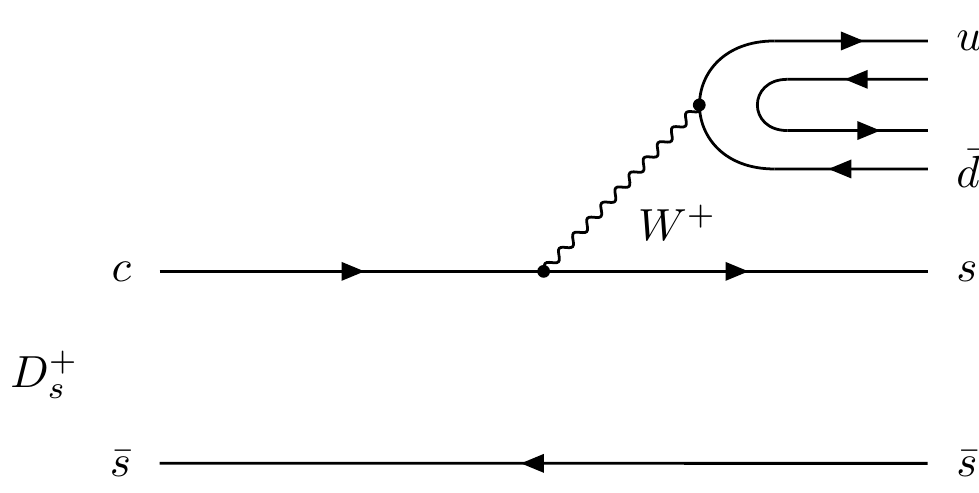} 
		\caption{\footnotesize $u\bar{d}$ quark pair hadronizes into a final meson pair.}
		\label{fig:Feynman2}  
	\end{subfigure}	
	\caption{Diagrams for the $D_{s}^{+} \rightarrow K_{S}^{0}K^{+}\pi^{0}$ decay with external $W$-emission mechanisms.}
	\label{fig:Feynman12}
\end{figure}

\begin{figure}[htbp]
	\begin{subfigure}{0.45\textwidth}
		\centering
		\includegraphics[width=1\linewidth,trim=150 530 180 120,clip]{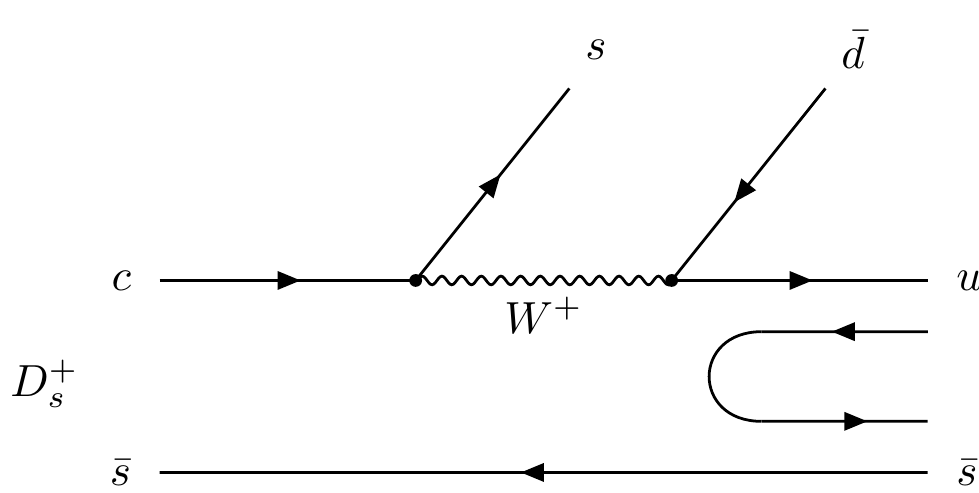} 
		\caption{\footnotesize $u\bar{s}$ quark pair hadronizes into a final meson pair.}
		\label{fig:Feynman3}
	\end{subfigure}
	\quad
	\quad
	\begin{subfigure}{0.45\textwidth}  
		\centering 
		\includegraphics[width=1\linewidth,trim=150 530 180 120,clip]{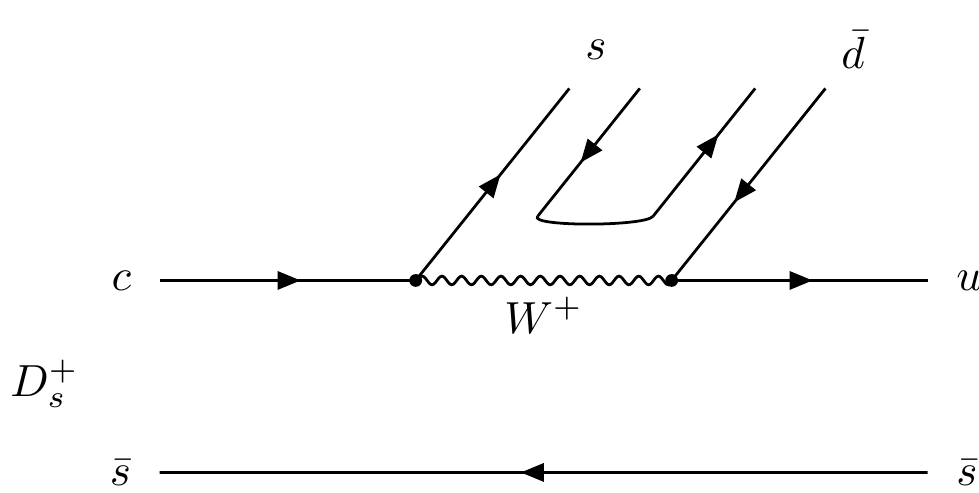} 
		\caption{\footnotesize $s\bar{d}$ quark pair hadronizes into a final meson pair.}
		\label{fig:Feynman4}  
	\end{subfigure}	
	\caption{Diagrams for the $D_{s}^{+} \rightarrow K_{S}^{0}K^{+}\pi^{0}$ decay with internal $W$-emission mechanisms.}
	\label{fig:Feynman34}
\end{figure}

For the three-body weak decay $D_{s}^{+} \rightarrow K_{S}^{0}K^{+}\pi^{0}$, we start from the dynamics at the quark level, where the dominant external and internal $W$-emission mechanisms \cite{Chau:1982da,Chau:1987tk} are taken into account. In the next step, in the hadron level we consider the final state interactions in the $S$-wave and the vector meson productions in the $P$-wave, which will be discussed later.
First, the Feynman diagrams of the external $W$-emission mechanisms are shown in Fig. \ref{fig:Feynman12}, and the ones with the internal $W$-emission mechanisms are given in Fig. \ref{fig:Feynman34}. As shown in Fig. \ref{fig:Feynman12} for the weak decays of $D_{s}^{+}$, the $c$ quark decays into a $W^{+}$ boson and an $s$ quark, and the $\bar{s}$ quark in $D_{s}^{+}$ as a spectator remains unchanged, then the $W^{+}$ boson decays into an $u\bar{d}$ quark pair. 
In the following procedures, there are two possibilities for the hadonization progresses. In Fig. \ref{fig:Feynman1}, the $u\bar{d}$ pair forms a $\pi^{+}$ or $\rho^{+}$ meson, along with this process, the $s\bar{s}$ quark pair hadronizes into two mesons with $\bar{q}q=\bar{u}u+\bar{d}d+\bar{s}s$ produced from the vacuum. 
Contrarily, in Fig. \ref{fig:Feynman2}, the $s\bar{s}$ quark pair goes into an $\eta$ or $\phi$ meson, the $u\bar{d}$ quark pair made by the $W^{+}$ boson hadronizes into two mesons with the $\bar{q}q$ pairs generated from the vacuum. The corresponding processes for these hadonizations can be given by the formulae below for Fig. \ref{fig:Feynman1} and Fig. \ref{fig:Feynman2}, respectively,
\begin{equation}
	\begin{aligned}
		|H^{(1a)}\rangle
		=&V_{P}^{(1a)}V_{cs}V_{ud}(u\bar{d}\rightarrow\pi^{+})|s(\bar{u}u+\bar{d}d+\bar{s}s)\bar{s}\rangle \\
		&+V_{P}^{*(1a)}V_{cs}V_{ud}(u\bar{d}\rightarrow\rho^{+})|s(\bar{u}u+\bar{d}d+\bar{s}s)\bar{s}\rangle \\
		=&V_{P}^{(1a)}V_{cs}V_{ud}\pi^{+}(M\cdot M)_{33}+V_{P}^{*(1a)}V_{cs}V_{ud}\rho^{+}(M\cdot M)_{33}, 
	\end{aligned}
	\label{eq:H1a}
\end{equation}
\begin{equation}
	\begin{aligned}
		|H^{(1b)}\rangle
		=&V_{P}^{(1b)}V_{cs}V_{ud}(s\bar{s}\rightarrow\frac{-2}{\sqrt{6}}\eta)|u(\bar{u}u+\bar{d}d+\bar{s}s)\bar{d}\rangle \\
		&+V_{P}^{*(1b)}V_{cs}V_{ud}(s\bar{s}\rightarrow\phi)|u(\bar{u}u+\bar{d}d+\bar{s}s)\bar{d}\rangle \\
		=&V_{P}^{(1b)}V_{cs}V_{ud}\frac{-2}{\sqrt{6}}\eta(M\cdot M)_{12}+V_{P}^{*(1b)}V_{cs}V_{ud}\phi(M\cdot M)_{12}, 
	\end{aligned}
	\label{eq:H1b}
\end{equation}
where the $V_{P}^{(1a)}$ and $V_{P}^{*(1a)}$ are the weak interaction strengths of the production vertices \cite{Liang:2015qva,Ahmed:2020qkv} for the generations $\pi^{+}$ and $\rho^{+}$, respectively, for the case of Fig. \ref{fig:Feynman1}, and the $V_{P}^{(1b)}$ and $V_{P}^{*(1b)}$ are the ones for the productions $\eta$ and $\phi$, severally, for the other case of Fig. \ref{fig:Feynman2}. 
The factors $V_{cs}$ and $V_{ud}$ are the elements of the Cabibbo-Kobayashi-Maskawa (CKM) matrix, which indicate from $q_{1}\rightarrow q_{2}$ quarks. 
The symbol $M$ is the $q\bar{q}$ matrix in $SU{(3)}$, defined as
\begin{equation}
	\begin{aligned}
		M=\left(\begin{array}{lll}{u \bar{u}} & {u \bar{d}} & {u \bar{s}} \\ {d \bar{u}} & {d \bar{d}} & {d \bar{s}} \\ {s \bar{u}} & {s \bar{d}} & {s \bar{s}}\end{array}\right).
	\end{aligned}
	\label{eq:M}
\end{equation}

\begin{figure}[tbp]
	\begin{subfigure}{0.45\textwidth}
		\centering
		\includegraphics[width=1\linewidth,trim=140 550 210 120,clip]{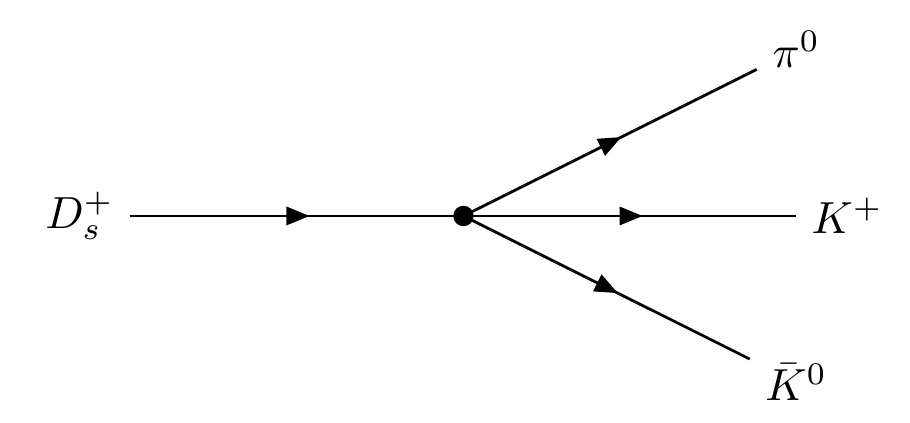} 
		\caption{\footnotesize Tree-level diagram.}
		\label{fig:Scatter1}
	\end{subfigure}
	\quad
	\quad
	\begin{subfigure}{0.45\textwidth}
		\centering
		\includegraphics[width=1\linewidth,trim=150 530 180 120,clip]{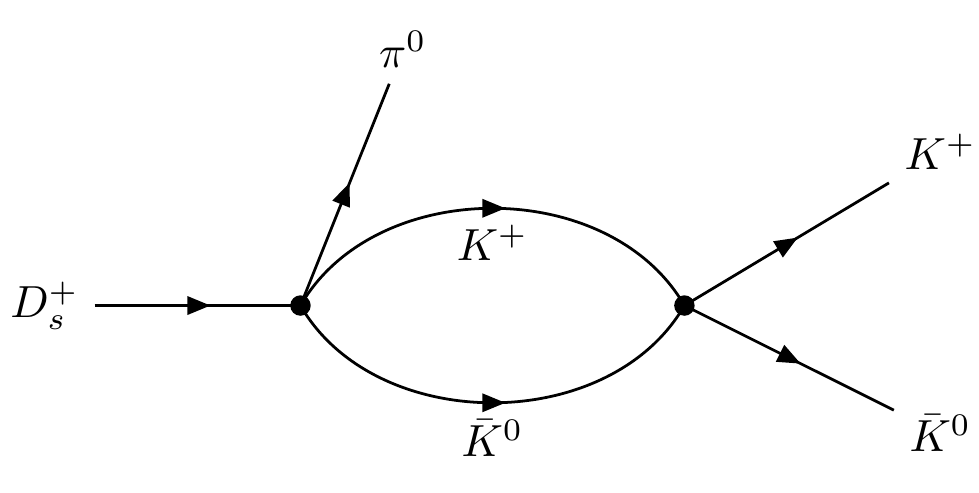} 
		\caption{\footnotesize Rescattering of the $K^{+}\bar{K}^{0}\rightarrow K^{+}\bar{K}^{0}$.}
		\label{fig:Scatter3}
	\end{subfigure}
	\quad
	\quad
	\begin{subfigure}{0.45\textwidth}
		\centering
		\includegraphics[width=1\linewidth,trim=150 530 180 120,clip]{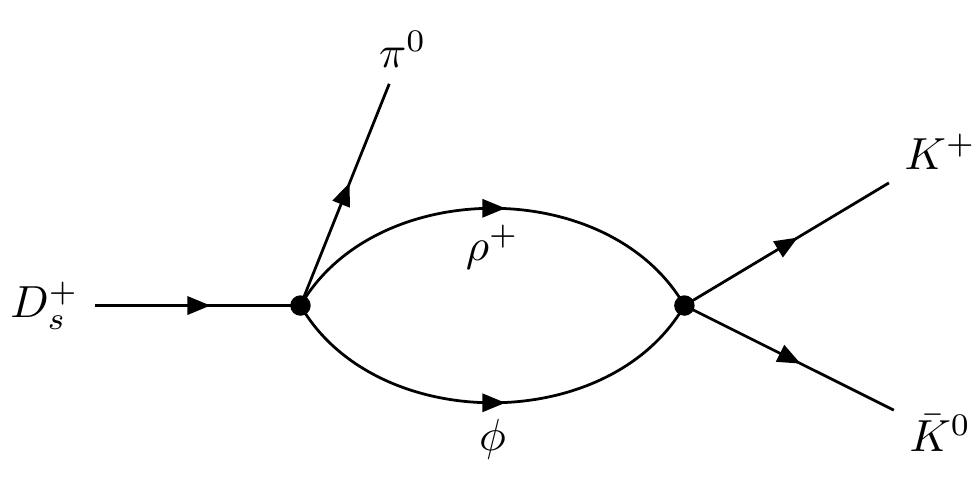} 
		\caption{\footnotesize Rescattering of the $\rho^{+}\phi \rightarrow K^{+}\bar{K}^{0}$.}
		\label{fig:Scatter2}
	\end{subfigure}
	\begin{subfigure}{0.45\textwidth}  
		\centering 
		\includegraphics[width=1\linewidth,trim=150 530 180 120,clip]{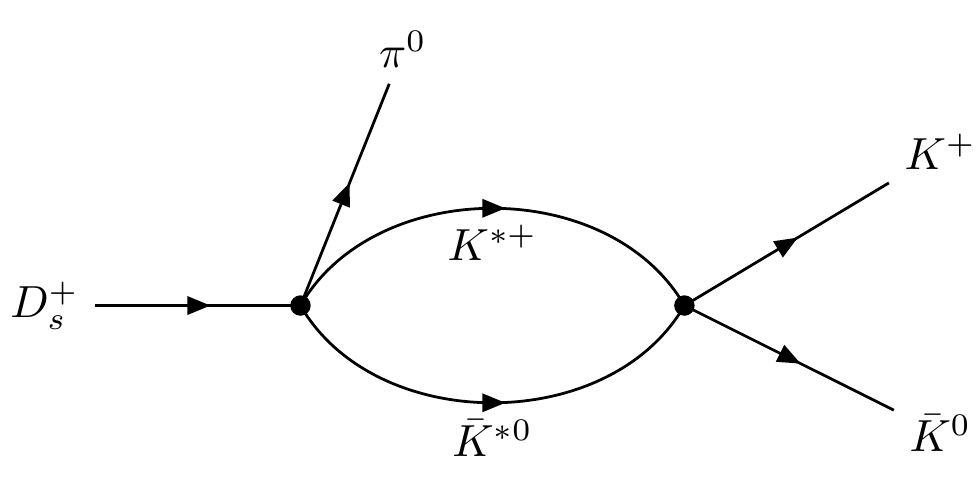} 
		\caption{\footnotesize Rescattering of the $K^{*+}\bar{K}^{*0}\rightarrow K^{+}\bar{K}^{0}$.}
		\label{fig:Scatter4}  
	\end{subfigure}	
	\caption{Mechanisms of the $S$-wave final state interactions in the $D_{s}^{+}$ decay.}
	\label{fig:Scatter}
\end{figure}

Analogously, in the mechanisms of internal $W$-emission, see in Fig. \ref{fig:Feynman34}, the $s\bar{d}$ pair goes into a $\bar{K}^{0}$ or $\bar{K}^{*0}$ meson, and the remnant $u\bar{s}$ quark pair hadronizes into two mesons with $\bar{q}q$ pairs produced from the vacuum, as shown in Fig. \ref{fig:Feynman3}. 
On the other hand, in Fig. \ref{fig:Feynman4}, the $u\bar{s}$ pair forms a $K^{+}$ or $K^{*+}$ meson, and the $s\bar{d}$ quark pair hadronizes into two mesons with $\bar{q}q$ pairs created from the vacuum. One can write these processes in the following way for Fig. \ref{fig:Feynman3} and Fig. \ref{fig:Feynman4}, respectively,
\begin{equation}
	\begin{aligned}
		|H^{(2a)}\rangle
		=&V_{P}^{(2a)}V_{cs}V_{ud}(s\bar{d}\rightarrow\bar{K}^{0})|u(\bar{u}u+\bar{d}d+\bar{s}s)\bar{s}\rangle \\
		&+V_{P}^{*(2a)}V_{cs}V_{ud}(s\bar{d}\rightarrow\bar{K}^{*0})|u(\bar{u}u+\bar{d}d+\bar{s}s)\bar{s}\rangle \\
		=&V_{P}^{(2a)}V_{cs}V_{ud}\bar{K}^{0}(M\cdot M)_{13}+V_{P}^{*(2a)}V_{cs}V_{ud}\bar{K}^{*0}(M\cdot M)_{13}, 
	\end{aligned}
	\label{eq:H2a}
\end{equation}
\begin{equation}
	\begin{aligned}
		|H^{(2b)}\rangle=&V_{P}^{(2b)}V_{cs}V_{ud}(u\bar{s}\rightarrow K^{+})|s(\bar{u}u+\bar{d}d+\bar{s}s)\bar{d}\rangle \\
		&+V_{P}^{*(2b)}V_{cs}V_{ud}(u\bar{s}\rightarrow K^{*+})|s(\bar{u}u+\bar{d}d+\bar{s}s)\bar{d}\rangle \\
		=&V_{P}^{(2b)}V_{cs}V_{ud}K^{+}(M\cdot M)_{32}+V_{P}^{*(2b)}V_{cs}V_{ud}K^{*+}(M\cdot M)_{32},
	\end{aligned}
	\label{eq:H2b}
\end{equation}
where the $V_{P}^{(2a)}$ and $V_{P}^{*(2a)}$ are the weak interaction strengths of the production vertices for the creations $\bar{K}^{0}$ and $\bar{K}^{*0}$, respectively, and the $V_{P}^{(2b)}$ and $V_{P}^{*(2b)}$ are the ones for the formations $K^{+}$ and $K^{*+}$, severally. 
Afterward, the matrix $M$ for the hadronization can be revised in terms of the pseudoscalar ($P$) or vector ($V$) mesons, written as
\begin{equation}
	\begin{aligned}
		P=\left(\begin{array}{ccc}{\frac{1}{\sqrt{2}} \pi^{0}+\frac{1}{\sqrt{6}} \eta} & {\pi^{+}} & {K^{+}} \\ {\pi^{-}} & {-\frac{1}{\sqrt{2}} \pi^{0}+\frac{1}{\sqrt{6}} \eta} & {K^{0}} \\ {K^{-}} & {\bar{K}^{0}} & {-\frac{2}{\sqrt{6}} \eta}\end{array}\right),
	\end{aligned}
\end{equation} 
\begin{equation}
	\begin{aligned}
		V=\left(\begin{array}{ccc}{\frac{1}{\sqrt{2}} \rho^{0}+\frac{1}{\sqrt{2}} \omega} & {\rho^{+}} & {K^{*+}} \\ {\rho^{-}} & {-\frac{1}{\sqrt{2}} \rho^{0}+\frac{1}{\sqrt{2}} \omega} & {K^{*0}} \\ {K^{*-}} & {\bar{K}^{*0}} & {\phi}\end{array}\right),
	\end{aligned}
\end{equation} 
where we take $\eta\equiv\eta_{8}$ \cite{Liang:2014tia}. In Eqs. (\ref{eq:H1a}), (\ref{eq:H1b}), (\ref{eq:H2a}), and (\ref{eq:H2b}), the $M\cdot M$ has four possible situations with two matrices of physical mesons, {\it i.e.}, $P\cdot P$, $V\cdot V$, $P\cdot V$, and $V\cdot P$. And thus, these hadronization processes can be reexpressed as
\begin{equation}
	\begin{aligned}
		|H^{(1b)}\rangle=V_{P}^{*(1b)}V_{cs}V_{ud}\phi(\frac{-1}{\sqrt{2}}\rho^{+}\pi^{0})+V_{P}^{*(1b)'}V_{cs}V_{ud}\phi(\frac{1}{\sqrt{2}}\rho^{+}\pi^{0}), 
	\end{aligned}
	\label{eq:H1b2}
\end{equation}
\begin{equation}
	\begin{aligned}
		|H^{(2a)}\rangle=V_{P}^{(2a)}V_{cs}V_{ud}\bar{K}^{0}(\frac{1}{\sqrt{2}}K^{+}\pi^{0})+V_{P}^{*(2a)}V_{cs}V_{ud}\bar{K}^{*0}(\frac{1}{\sqrt{2}}K^{*+}\pi^{0}), 
	\end{aligned}
	\label{eq:H2a2}
\end{equation}
\begin{equation}
	\begin{aligned}
		|H^{(2b)}\rangle=V_{P}^{(2b)}V_{cs}V_{ud}K^{+}(-\frac{1}{\sqrt{2}}\bar{K}^{0}\pi^{0})+V_{P}^{*(2b)}V_{cs}V_{ud}K^{*+}(-\frac{1}{\sqrt{2}}\bar{K}^{*0}\pi^{0}), 
	\end{aligned}
	\label{eq:H2b2}
\end{equation}
where we only keep the terms that contribute to the final states $K_{S}^{0}K^{+}\pi^{0}$. It should be mentioned that there is no term for the Fig. \ref{fig:Feynman1} contributed to these final states of present $D_{s}^{+}$ decay process. Besides, in Eq. (\ref{eq:H1b2}), the factors $V_{P}^{*(1b)}$ and $V_{P}^{*(1b)'}$ are different because they come from $(V\cdot P)_{12}$ and $(P\cdot V)_{12}$, respectively. Then, we obtain the total contributions in the $S$-wave,
\begin{equation}
	\begin{aligned}
		|H\rangle=&|H^{(1b)}\rangle+|H^{(2a)}\rangle+|H^{(2b)}\rangle \\
		=&\frac{1}{\sqrt{2}}(V_{P}^{*(1b)'}-V_{P}^{*(1b)})V_{cs}V_{ud}\rho^{+}\phi\pi^{0}+\frac{1}{\sqrt{2}}(V_{P}^{(2a)}-V_{P}^{(2b)})V_{cs}V_{ud}K^{+}\bar{K}^{0}\pi^{0} \\
		&+\frac{1}{\sqrt{2}}(V_{P}^{*(2a)}-V_{P}^{*(2b)})V_{cs}V_{ud}K^{*+}\bar{K}^{*0}\pi^{0}, \\
		=&\frac{1}{\sqrt{2}}V_{P}^{*'}V_{cs}V_{ud}\rho^{+}\phi\pi^{0}
		+\frac{1}{\sqrt{2}}V_{P}V_{cs}V_{ud}K^{+}\bar{K}^{0}\pi^{0}
		+\frac{1}{\sqrt{2}}V_{P}^{*}V_{cs}V_{ud}K^{*+}\bar{K}^{*0}\pi^{0} , 
	\end{aligned}
	\label{eq:H}
\end{equation}
where we define $V_{P}^{*'}=V_{P}^{*(1b)'}-V_{P}^{*(1b)}$, $V_{P}=V_{P}^{(2a)}-V_{P}^{(2b)}$, and $V_{P}^{*}=V_{P}^{*(2a)}-V_{P}^{*(2b)}$. Note that there are also the final states $\bar{K}^{0}K^{+}\pi^{0}$ produced directly in the hadronization processes in Eq. (\ref{eq:H}). Taking into account the final state interactions, we can get these final states via the rescattering procedures, such as $K^{+}\bar{K}^{0} \rightarrow K^{+}\bar{K}^{0}$, $\rho^{+}\phi \rightarrow K^{+}\bar{K}^{0}$, and $K^{*+}\bar{K}^{*0} \rightarrow K^{+}\bar{K}^{0}$, which are depicted in Fig. \ref{fig:Scatter}. 
One more thing should be mentioned that there is no direct term $\eta\pi^+ \pi^0$ contributed in Eq. (\ref{eq:H}) due to its two terms in $|H^{(1b)}\rangle$ cancelled, which is consistent with the evaluation of Ref. \cite{Molina:2019udw}, where the experimental findings for the decay $D^+_s \to \eta\pi^+ \pi^0$ \cite{BESIII:2019jjr} were investigated. In Ref. \cite{Molina:2019udw}, the larger decay rate of $D^+_s \to \eta\pi^+ \pi^0$ was explained via the internal $W$-emission mechanism for the decay process rather than the $W$-annihilation procedure as assumed in Ref. \cite{BESIII:2019jjr}. In the further study of the decay $D^+_s \to \eta\pi^+ \pi^0$, no tree diagram of $D^+_s \to \eta\pi^+ \pi^0$ decay was taken into account in Refs. \cite{Hsiao:2019ait,Ling:2021qzl}. 
Therefore, under the dominant external and internal $W$-emission mechanisms, the amplitude of decay $D_{s}^{+} \rightarrow \bar{K}^{0}K^{+}\pi^{0}$ in the $S$-wave is given by
\begin{equation}
	\begin{aligned} 
		t_{S\text{-wave}}(M_{12}) \big|_{\bar{K}^{0}K^{+}\pi^{0}}
		=&\frac{1}{\sqrt{2}}\mathcal{C}_{1}G_{\rho^{+}\phi}(M_{12}) 
		T_{\rho^{+}\phi \rightarrow K^{+}\bar{K}^{0}}(M_{12}) \\
		&+\frac{1}{\sqrt{2}}\mathcal{C}_{2} 
		+\frac{1}{\sqrt{2}}\mathcal{C}_{2}G_{K^{+}\bar{K}^{0}}(M_{12}) 
		T_{K^{+}\bar{K}^{0} \rightarrow K^{+}\bar{K}^{0}}(M_{12}) \\
		&+\frac{1}{\sqrt{2}}\mathcal{C}_{3}G_{K^{*+}\bar{K}^{*0}}(M_{12}) 
		T_{K^{*+}\bar{K}^{*0} \rightarrow K^{+}\bar{K}^{0}}(M_{12}),
	\end{aligned}
	\label{eq:amplitudes}
\end{equation}
where the factors $V_{P}^{*'}V_{cs}V_{ud}$, $V_{P}V_{cs}V_{ud}$, and $V_{P}^{*}V_{cs}V_{ud}$ in Eq. (\ref{eq:H}) have been absorbed into the parameters $\mathcal{C}_{1}$, $\mathcal{C}_{2}$, and $\mathcal{C}_{3}$, respectively. In the present work, we take them as the free constants, which are independent on the invariant masses and contain the global normalization factor for matching the events of the experimental data. $M_{ij}$ is the energy of two particles in the center-of-mass (c.m.) frame, where the lower indices $i,j=1,2,3$ denote the three final states of $K_{S}^{0}(\bar{K}^{0})$, $K^{+}$, and $\pi^{0}$, respectively. Besides, $G_{PP^{'}(VV^{'})}$ and $T_{PP^{'}(VV^{'}) \rightarrow PP^{'}}$ are the loop functions and the two-body scattering amplitudes, respectively. Then, as done in Ref. \cite{Dai:2021owu}, we take $|K_{S}^{0}\rangle=\frac{1}{\sqrt{2}}(|K^{0}\rangle-|\bar{K}^{0}\rangle)$,  and change the final state from $\bar{K}^{0}$ to $K_{S}^{0}$, where Eq. (\ref{eq:amplitudes}) becomes
\begin{equation}
	\begin{aligned} 
		t_{S\text{-wave}}(M_{12}) \big|_{K^{0}_S K^{+}\pi^{0}}
		=&-\frac{1}{2}\mathcal{C}_{1}G_{\rho^{+}\phi}(M_{12}) 
		T_{\rho^{+}\phi \rightarrow K^{+}\bar{K}^{0}}(M_{12}) \\
		&-\frac{1}{2}\mathcal{C}_{2} 
		-\frac{1}{2}\mathcal{C}_{2}G_{K^{+}\bar{K}^{0}}(M_{12}) 
		T_{K^{+}\bar{K}^{0} \rightarrow K^{+}\bar{K}^{0}}(M_{12}) \\
		&-\frac{1}{2}\mathcal{C}_{3}G_{K^{*+}\bar{K}^{*0}}(M_{12}) 
		T_{K^{*+}\bar{K}^{*0} \rightarrow K^{+}\bar{K}^{0}}(M_{12}).
	\end{aligned}
	\label{eq:amplitudes2}
\end{equation}

\begin{figure}[htbp]
	\begin{subfigure}{0.45\textwidth}
		\centering
		\includegraphics[width=1\linewidth,trim=150 580 260 120,clip]{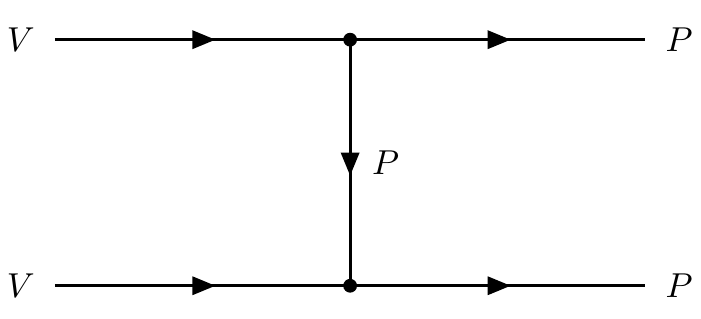} 
		\caption{\footnotesize t-channel.}
		\label{fig:tchannel}
	\end{subfigure}
	\quad
	\quad
	\begin{subfigure}{0.45\textwidth}  
		\centering 
		\includegraphics[width=1\linewidth,trim=150 580 260 120,clip]{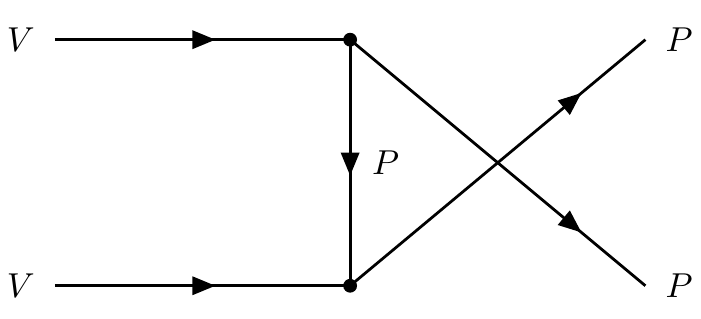} 
		\caption{\footnotesize u-channel.}
		\label{fig:uchannel}  
	\end{subfigure}	
	\caption{Feynman diagrams of t- and u-channels.}
	\label{fig:tandu}
\end{figure}

Furthermore, the rescattering amplitudes $T_{\rho^{+}\phi \rightarrow K^{+}\bar{K}^{0}}$, $T_{K^{+}\bar{K}^{0} \rightarrow K^{+}\bar{K}^{0}}$, and $T_{K^{*+}\bar{K}^{*0} \rightarrow K^{+}\bar{K}^{0}}$ in Eq. (\ref{eq:amplitudes2}) can be calculated  by the coupled channel Bethe-Salpeter equation of the on-shell form,
\begin{equation}
	\begin{aligned} 
		T = [1-vG]^{-1}v, 
	\end{aligned}
	\label{eq:BSE}
\end{equation}
where the matrix $v$ is constituted by the $S$-wave interaction potentials in the coupled channels. In the present work, we consider the interactions of five channels $K^{*+}\bar{K}^{*0}$, $\rho^{+}\omega$, $\rho^{+}\phi$, $K^{+}\bar{K}^{0}$, and $\pi^{+}\eta$, where one can expect that the states $a_{0}(980)$ and $a_{0}(1710)$ will be dynamically generated. Among them, the potential elements of $v_{VV\rightarrow VV}$ are taken from the Appendix of Ref. \cite{Geng:2008gx} (the arXiv version), which included the contact and exchange vector meson terms. The $VVP$ vertex is suppressed, and thus, the contributions of exchange pseudoscalar meson are ignored. As done in Ref. \cite{Wang:2019niy}, the potentials $v_{PP\rightarrow PP}$ are taken from Refs. \cite{Oller:1997ti,Duan:2020vye,Xie:2014tma}, which only included the contact items from the chiral Lagrangian. The ones $v_{VV\rightarrow PP}$ are evaluated with the approach of Ref. \cite{Wang:2022pin}, where the Feynman diagrams of t- and u-channels were considered, as depicted in Fig. \ref{fig:tandu}. The interaction Lagrangian for the $VPP$ vertex is given by \cite{Bando:1984ej,Bando:1987br},
\begin{equation}
	\begin{aligned} 
		\mathcal{L}_{V P P}=-i g\left\langle V_\mu\left[P, \partial^\mu P\right]\right\rangle,
	\end{aligned}
	\label{eq:Lagrangian}
\end{equation}
with $g=M_{V}/(2f_{\pi})$, where taking $M_{V}=0.84566$ GeV is the averaged vector-meson mass and $f_{\pi}=0.093$ GeV pion decay constant, which are taken from Ref. \cite{Wang:2022pin}. Thus, the interaction potentials are given by
\begin{equation}
	\begin{aligned} 
		&v_{K^{*+}\bar{K}^{*0}\rightarrow K^{+}\bar{K}^{0}}=\left(\frac{2}{t-m_{\pi}^{2}}-\frac{6}{t-m_{\eta}^{2}}\right)
		g^{2}\epsilon_{1\mu}k_{3}^{\mu}\epsilon_{2\nu}k_{4}^{\nu}, \\
		&v_{K^{*+}\bar{K}^{*0}\rightarrow\pi^{+}\eta}=-2\sqrt{6}\left(\frac{g^{2}}{t-m_{K}^{2}}
		\epsilon_{1\mu}k_{3}^{\mu}\epsilon_{2\nu}k_{4}^{\nu}+\frac{g^{2}}{u-m_{K}^{2}}
		\epsilon_{1\mu}k_{4}^{\mu}\epsilon_{2\nu}k_{3}^{\nu}\right), \\
		&v_{\rho^{+}\omega\rightarrow K^{+}\bar{K}^{0}}=-2\sqrt{2}\left(\frac{g^{2}}{t-m_{K}^{2}}
		\epsilon_{1\mu}k_{3}^{\mu}\epsilon_{2\nu}k_{4}^{\nu}+\frac{g^{2}}{u-m_{K}^{2}}
		\epsilon_{1\mu}k_{4}^{\mu}\epsilon_{2\nu}k_{3}^{\nu}\right), \\
		&v_{\rho^{+}\omega\rightarrow\pi^{+}\eta}=0, \\
		&v_{\rho^{+}\phi\rightarrow K^{+}\bar{K}^{0}}=4\left(\frac{g^{2}}{t-m_{K}^{2}}
		\epsilon_{1\mu}k_{3}^{\mu}\epsilon_{2\nu}k_{4}^{\nu}+\frac{g^{2}}{u-m_{K}^{2}}
		\epsilon_{1\mu}k_{4}^{\mu}\epsilon_{2\nu}k_{3}^{\nu}\right), \\
		&v_{\rho^{+}\phi\rightarrow\pi^{+}\eta}=0,
	\end{aligned}
	\label{eq:V}
\end{equation}
where $t=(k_{1}-k_{3})^{2}$ and $u=(k_{1}-k_{4})^{2}$ are defined, $\epsilon_{i}$ is polarization vector, and $k_{i}$ three-momentum of the corresponding particles with the lower index $i$ ($i=1,2,3,4$) denoting the particles in scattering process $V(1)V(2)\rightarrow P(3)P(4)$. Compared with $v_{VV\rightarrow VV}$, the potentials of $v_{VV\rightarrow PP}$ are much strengthened, and thus, a monopole form factor is introduced at each $VPP$ vertex of the exchanged pseudoscalar meson as done in Refs. \cite{Molina:2008jw,Oset:2012zza,Wang:2021jub},
\begin{equation}
	\begin{aligned} 
		F=\frac{\Lambda^2-m_{e x}^2}{\Lambda^2-q^2},
	\end{aligned}
	\label{eq:FF}
\end{equation}
where $m_{e x}$ is the mass of the exchanged pseudoscalar meson, and $q$ the transferred momentum. The value of parameter $\Lambda$ is empirically chosen as $1.0$ GeV. After performing the partial wave projection, one can obtain the $S$-wave potentials $v$. 

The diagonal matrix $G$ is made up of the meson-meson two-point loop functions, where the explicit form of the element of matrix $G$ with the dimensional regularization is given by \cite{Oller:2000fj,Oller:1998zr,Gamermann:2006nm,Alvarez-Ruso:2010rqm,Guo:2016zep},
\begin{equation}
	\begin{aligned}
		G_{ii}(M_{inv})=& \frac{1}{16 \pi^{2}}\left\{a_{ii}(\mu)+\ln \frac{m_{1}^{2}}{\mu^{2}}+\frac{m_{2}^{2}-m_{1}^{2}+M_{inv}^{2}}{2M_{inv}^{2}} \ln \frac{m_{2}^{2}}{m_{1}^{2}}\right.\\
		&+\frac{q_{cmi}(M_{inv})}{M_{inv}}\left[\ln \left(M_{inv}^{2}-\left(m_{2}^{2}-m_{1}^{2}\right)+2 q_{cmi}(M_{inv}) M_{inv}\right)\right.\\
		&+\ln \left(M_{inv}^{2}+\left(m_{2}^{2}-m_{1}^{2}\right)+2 q_{cmi}(M_{inv})M_{inv}\right) \\
		&-\ln \left(-M_{inv}^{2}-\left(m_{2}^{2}-m_{1}^{2}\right)+2 q_{cmi}(M_{inv})M_{inv}\right) \\
		&\left.\left.-\ln \left(-M_{inv}^{2}+\left(m_{2}^{2}-m_{1}^{2}\right)+2 q_{cmi}(M_{inv})M_{inv}\right)\right]\right\},
	\end{aligned}
	\label{eq:DR}
\end{equation}
where $m_{1}$ and $m_{2}$ are the masses of the intermediate mesons in the loops, $M_{inv}$ is the invariant mass of the meson-meson system, $\mu$ the regularization scale, of which the value will be discussed in Sec. \ref{sec:Results}, and the $a_{ii}(\mu)$ the subtraction constant. As done in Refs. \cite{Duan:2020vye,Wang:2021kka}, its value can be evaluated by Eq. (17) of Ref. \cite{Oller:2000fj},
\begin{equation}
	\begin{aligned}
		a_{ii}(\mu)=-2 \ln \left(1+\sqrt{1+\frac{m_{1}^{2}}{\mu^{2}}}\right)+\cdots,
	\end{aligned}
	\label{eq:aii}
\end{equation}
where $m_{1}$ is the mass of a larger-mass meson in the corresponding channels, and the ellipses indicates the ingnored higher order terms in the nonrelativistic expansion \cite{Guo:2018tjx}. Besides, $q_{cmi}(M_{inv})$ is the three-momentum of the particle in the c.m. frame,
\begin{equation}
	\begin{aligned}
		q_{cmi}(M_{inv})=\frac{\lambda^{1 / 2}\left(M_{inv}^{2}, m_{1}^{2}, m_{2}^{2}\right)}{2M_{inv}},
	\end{aligned}
	\label{eq:qcm}
\end{equation}
with the usual K\"allen triangle function $\lambda(a, b, c)=a^{2}+b^{2}+c^{2}-2(a b+a c+b c)$.

\begin{figure}[tbp]
	\begin{subfigure}{0.45\textwidth}
		\centering
		\includegraphics[width=1\linewidth,trim=150 540 180 120,clip]{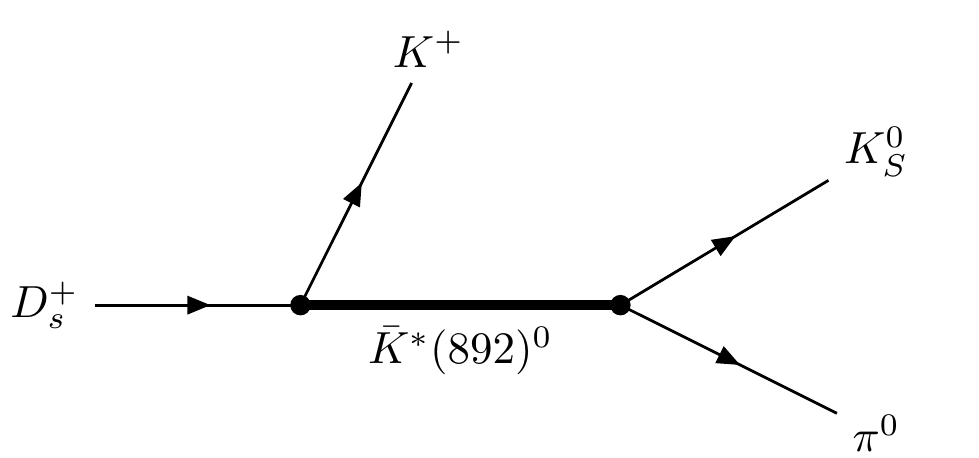} 
		\caption{\footnotesize Diagram via the $\bar{K}^{*}(892)^{0}$.}
		\label{fig:Kstar892Z}
	\end{subfigure}
	\quad
	\quad
	\begin{subfigure}{0.45\textwidth}  
		\centering 
		\includegraphics[width=1\linewidth,trim=150 540 180 120,clip]{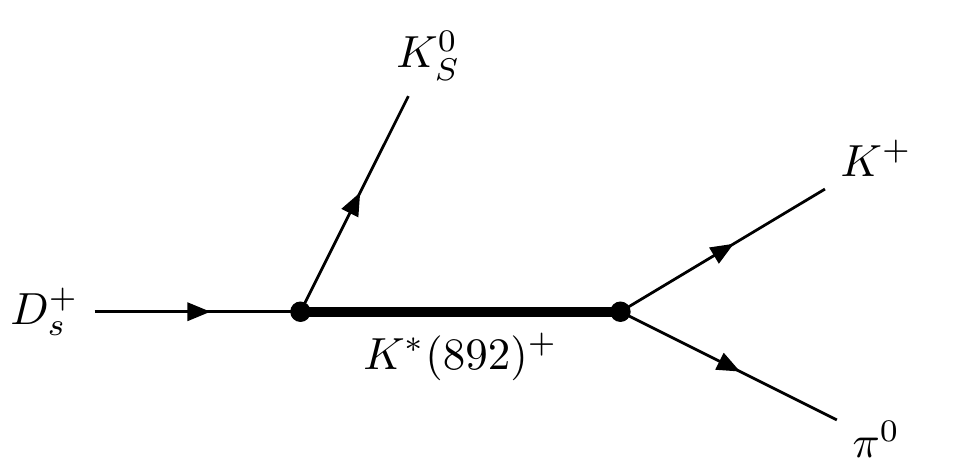} 
		\caption{\footnotesize Diagram via the $K^{*}(892)^{+}$.}
		\label{fig:Kstar892P}  
	\end{subfigure}	
	\caption{Mechanisms of $D_{s}^{+} \rightarrow K_{S}^{0}K^{+}\pi^{0}$ decay via the intermediate states $\bar{K}^{*}(892)^{0}$ and $K^{*}(892)^{+}$.}
	\label{fig:Kstar892}
\end{figure}

In addition, we also consider the contributions of the vector resonances in the intermediate states in the $P$-wave as discussed above, such as the ones $\bar{K}^{*}(892)^{0}$ and $K^{*}(892)^{+}$, which are not produced in the meson-meson scattering amplitudes. The production mechanisms are depicted in Fig. \ref{fig:Kstar892}. Referring to Refs. \cite{Toledo:2020zxj,Roca:2020lyi}, the relativistic amplitude for the decay $D_{s}^{+} \rightarrow \bar{K}^{*}(892)^{0}K^{+}\rightarrow K_{S}^{0}\pi^{0}K^{+}$ can be written as,
\begin{equation}
	\begin{aligned}
		t_{\bar{K}^{*}(892)^{0}}(M_{12},M_{13})=& \frac{\mathcal D_{1}e^{i\phi_{\bar{K}^{*}(892)^{0}}}} {M_{13}^{2}-m_{\bar{K}^{*}(892)^{0}}^{2}+i m_{\bar{K}^{*}(892)^{0}} \Gamma_{\bar{K}^{*}(892)^{0}}} \\
		&\times \left[(m_{K_{S}^{0}}^{2}-m_{\pi^{0}}^{2})
		\frac{m_{D_{s}^{+}}^{2}-m_{K^{+}}^{2}}{m_{\bar{K}^{*}(892)^{0}}^{2}}-M_{12}^{2}+M_{23}^{2}\right],
	\end{aligned}
	\label{eq:amplitudeKstarZ}
\end{equation}
where $\mathcal D_{1}$ is a unknown constant, $\phi_{\bar{K}^{*}(892)^{0}}$ a phase for the interference effect, the mass of $\bar{K}^{*}(892)^{0}$ taken as $m_{\bar{K}^{*}(892)^{0}}=0.89555$ GeV, and the width taken as $\Gamma_{\bar{K}^{*}(892)^{0}}=0.0473$ GeV, both of which
are taken from the Particle Data Group (PDG) \cite{Workman:2022ynf}. Note that the invariant masses $M_{ij}$ fulfil the constraint condition,
\begin{equation}
	\begin{aligned}
		M_{12}^{2}+M_{13}^{2}+M_{23}^{2}=m_{D_{s}^{+}}^{2}+m_{K_{S}^{0}}^{2}+m_{K^{+}}^{2}+m_{\pi^{0}}^{2}.
	\end{aligned}
	\label{eq:Mij}
\end{equation}

Analogously, the amplitude for the decay $D_{s}^{+} \rightarrow K_{S}^{0}K^{*}(892)^{+}\rightarrow K_{S}^{0}K^{+}\pi^{0}$ is given by
\begin{equation}
	\begin{aligned}
		t_{K^{*}(892)^{+}}(M_{12},M_{13})=& \frac{\mathcal D_{2}e^{i\phi_{K^{*}(892)^{+}}}} {M_{23}^{2}-m_{K^{*}(892)^{+}}^{2}+i m_{K^{*}(892)^{+}} \Gamma_{K^{*}(892)^{+}}} \\
		&\times \left[(m_{K^{+}}^{2}-m_{\pi^{0}}^{2})
		\frac{m_{D_{s}^{+}}^{2}-m_{K_{S}^{0}}^{2}}{m_{K^{*}(892)^{+}}^{2}}-M_{12}^{2}+M_{13}^{2}\right],
	\end{aligned}
	\label{eq:amplitudeKstarP}
\end{equation}
where $\mathcal D_{2}$ is also a unknown constant, $\phi_{K^{*}(892)^{+}}$ a phase, the mass of $K^{*}(892)^{+}$ taken as $m_{K^{*}(892)^{+}}=0.89167$ GeV, and the width taken as $\Gamma_{K^{*}(892)^{+}}=0.0514$ GeV \cite{Workman:2022ynf}. 

Finally, according to the formula of Ref. \cite{Workman:2022ynf},  the double differential width of the decay $D_{s}^{+} \rightarrow K_{S}^{0}K^{+}\pi^{0}$ is obtained as
\begin{equation}
	\begin{aligned}
		\frac{d^{2} \Gamma}{d M_{12}d M_{13}}=\frac{1}{(2 \pi)^{3}} \frac{M_{12}M_{13}}{8 m_{D_{s}^{+}}^{3}} \left(\left|t_{S\text{-wave}} + t_{\bar{K}^{*}(892)^{0}}+t_{K^{*}(892)^{+}}\right|^{2}\right),
	\end{aligned}
	\label{eq:dGamma}
\end{equation}
where we have considered the interference between the $S$- and $P$-waves with a coherent sum for the amplitudes. Even though the scattering amplitudes of Eq. \eqref{eq:amplitudes2} are pure $S$-wave contribution in our formalism, the amplitudes in $P$-wave, see Eqs. \eqref{eq:amplitudeKstarZ} and \eqref{eq:amplitudeKstarP}, are in fact the Breit-Wigner type, which are not pure $P$-wave, and thus lead to nonzero interference with the $S$-wave amplitudes
\footnote{Thanks the referee for the useful comment. Indeed, without the interference effect, the contribution of the $a_0(1710)$ resonance will be enhanced, where our conclusions would not be changed.}. 
Note that, in the experimental modelling, the nonzero unphysical interference is always taken into account when the Breit-Wigner type amplitudes are used for the resonances.
With Eq. \eqref{eq:dGamma}, it is easy to calculate $d\Gamma/dM_{K_{S}^{0}K^{+}}$, $d\Gamma/dM_{K_{S}^{0}\pi^{0}}$, and $d\Gamma/dM_{K^{+}\pi^{0}}$ by integrating over each of the invariant mass variables with the limits of the Dalitz Plot, see Ref. \cite{Workman:2022ynf} for more details.

\section{Results}
\label{sec:Results}

\begin{figure}[htbp]
	\begin{subfigure}{0.45\textwidth}
		\centering
		\includegraphics[width=1\linewidth]{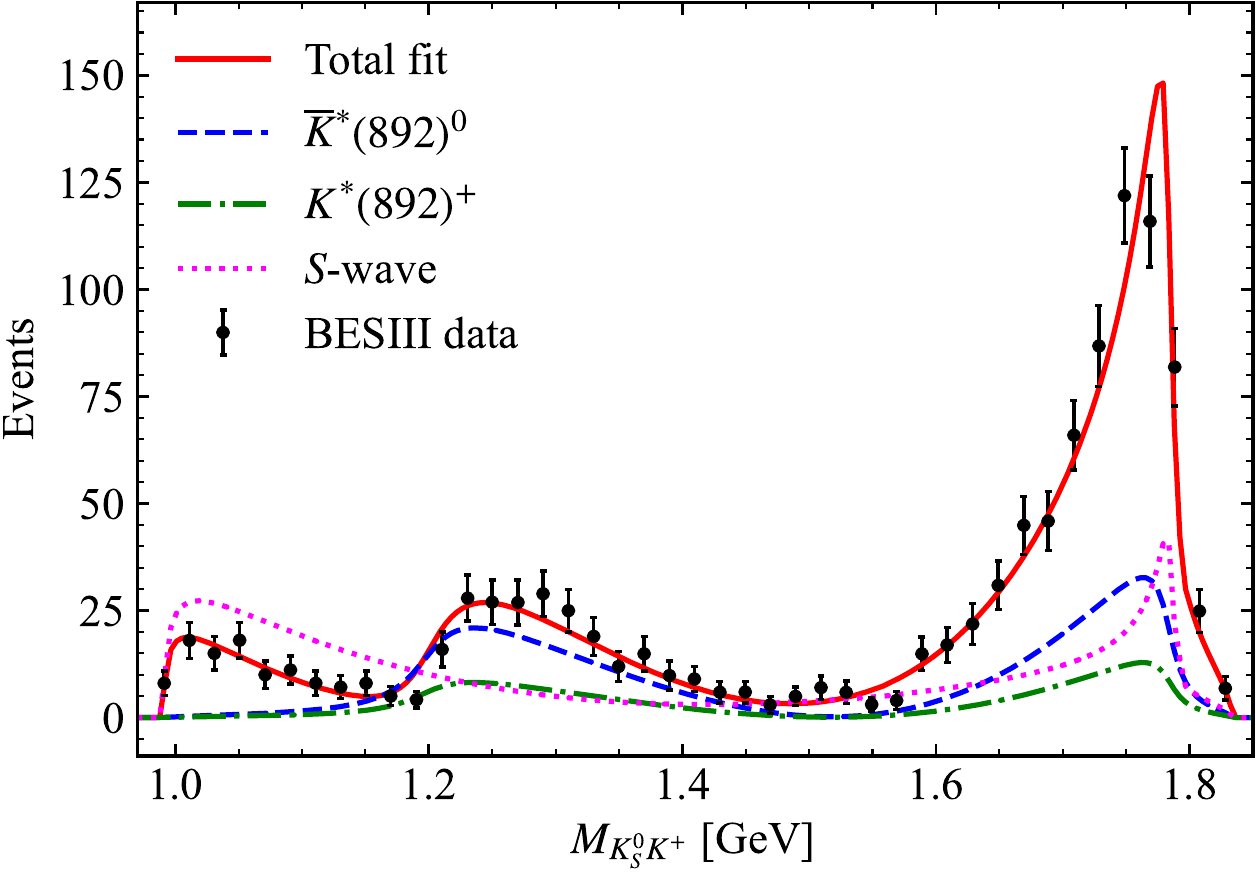} 
		\caption{\footnotesize $K_{S}^{0}K^{+}$ invariant mass distribution.}
		\label{fig:Fig12}
	\end{subfigure}
	\quad
	\quad
	\begin{subfigure}{0.45\textwidth}  
		\centering 
		\includegraphics[width=1\linewidth]{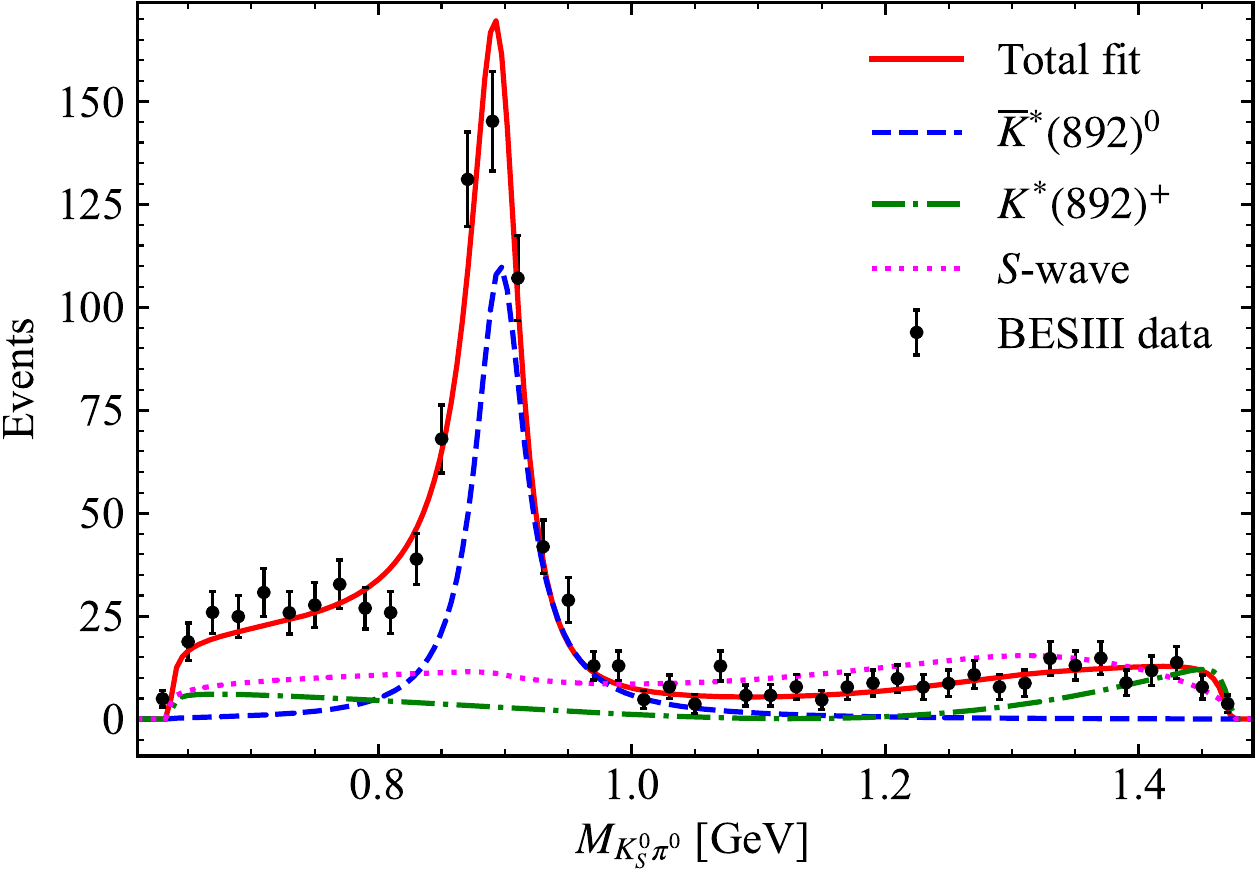} 
		\caption{\footnotesize $K_{S}^{0}\pi^{0}$ invariant mass distribution.}
		\label{fig:Fig13}  
	\end{subfigure}	
	\begin{subfigure}{0.45\textwidth}  
	\centering 
	\includegraphics[width=1\linewidth]{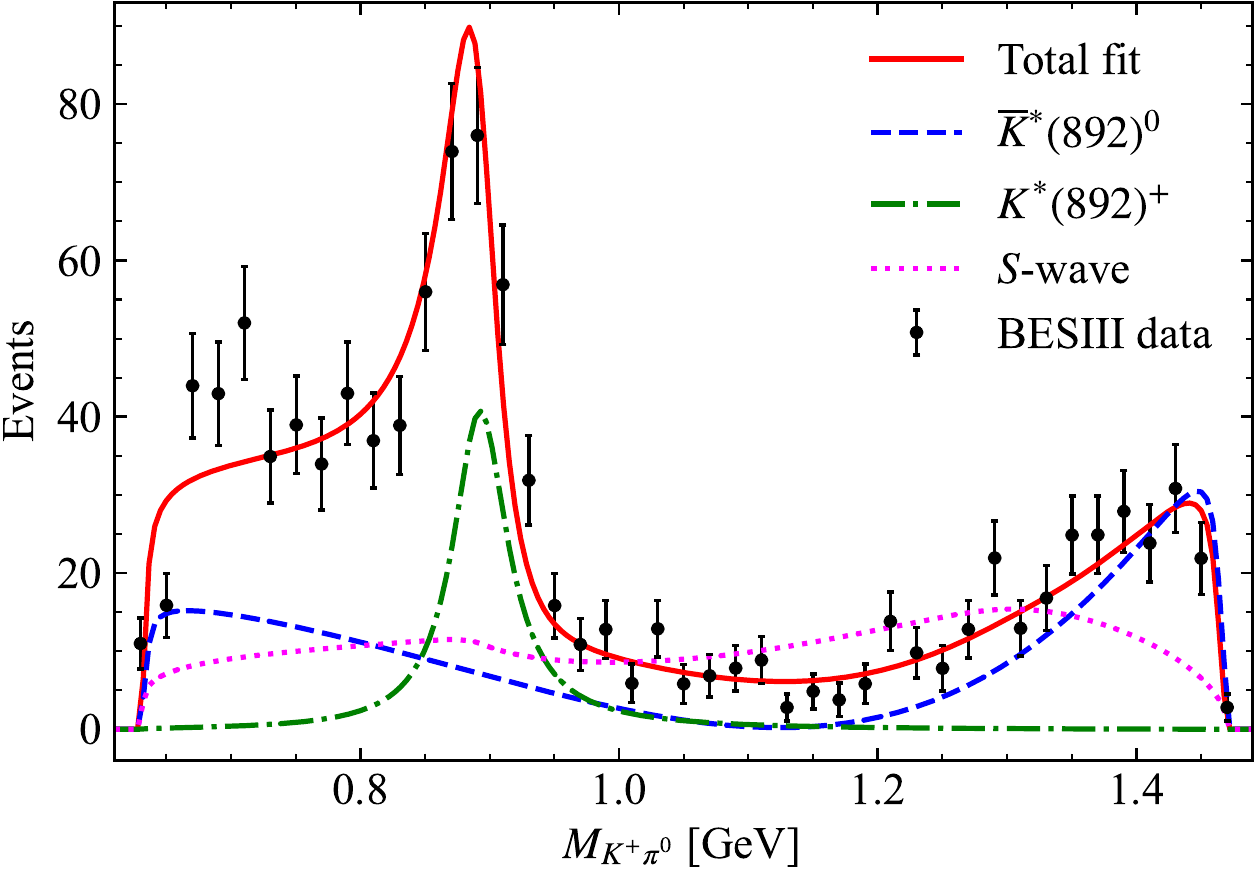} 
	\caption{\footnotesize $K^{+}\pi^{0}$ invariant mass distribution.}
	\label{fig:Fig23}  
\end{subfigure}	
	\caption{Invariant mass distributions for the $D_{s}^{+} \rightarrow K_{S}^{0}K^{+}\pi^{0}$ decay. The solid (red) line corresponds to the total contributions of the $S$- and $P$-waves, the dash (blue) line represents the contributions from the $\bar{K}^{*}(892)^{0}$, the dash-dot (green) line is the contributions from the $K^{*}(892)^{+}$, the dot (magenta) line is the contributions from the $S$-wave interactions ($a_{0}(980)^{+}$ and $a_{0}(1710)^{+}$), and the dot (black) points are the data taken from Ref. \cite{BESIII:2022npc}.}
	\label{fig:Fig}
\end{figure}

As one can see in the last section of our theoretical model, that we have eight parameters: the $\mu$ is the regularization scale in loop functions, $\mathcal C_{1}$, $\mathcal C_{2}$, and $\mathcal C_{3}$ represent the strengths of three production factors in the $S$-wave final state interactions, $\mathcal D_{1}$, $\mathcal D_{2}$, $\phi_{\bar{K}^{*}(892)^{0}}$, and $\phi_{K^{*}(892)^{+}}$ are the production factors and phases appeared in the $P$-wave productions, respectively. 
We perform a combined fit to the three invariant mass distributions of the $D_{s}^{+} \rightarrow K_{S}^{0}K^{+}\pi^{0}$ decay measured by the BESIII Collaboration \cite{BESIII:2022npc}. For the regularization scale $\mu$ in the loop functions of Eq. (\ref{eq:DR}), generally, the values $\mu=0.6$ GeV \cite{Duan:2020vye,Wang:2021ews} and $\mu=1.0$ GeV \cite{Geng:2008gx} are adopted in pseudoscalar-pseudoscalar and vector-vector meson interactions, respectively. In our fitting, we take it as a free parameter because our model includes the interactions with both pseudoscalar and vector meson channels 
\footnote{When we fix $\mu=1.0$ GeV, the fitting results are just a bit worse with $\chi^{2}/dof.=201.70/(129-7)=1.65$.}.
The parameters obtained from the fit are given in Table \ref{tab:Parameters}, where the fitted $\chi^{2}/dof.=171.24/(129-8)=1.42$, and the corresponding three invariant mass distributions are shown in Fig. \ref{fig:Fig}. When the regularization scale $\mu$ is taken as $0.716$ GeV from the fit, see Table \ref{tab:Parameters}, the subtraction constants $a_{ii}(\mu)$ for each coupled channels calculated by Eq. (\ref{eq:aii}) are obtained as
\begin{equation}
	\begin{aligned}
		&a_{K^{*+}\bar{K}^{*0}}=-1.91, \quad a_{\rho^{+}\omega}=-1.82, \quad a_{\rho^{+}\phi}=-2.02, \\
		&a_{K^{+}\bar{K}^{0}}=-1.59, \quad a_{\pi^{+}\eta}=-1.63.
	\end{aligned}
	\label{eq:amu}
\end{equation}
It should be mentioned that the uncertainties of the experimental data near the peak structures are larger than the others \cite{BESIII:2022npc}, as one can see in Fig. \ref{fig:Fig}. But, when we ignore the errors of the data, or equivalently set all the errors as $1$, the obtained results are not much different with the ones as shown in Fig. \ref{fig:Fig} with $\chi^{2}/dof.=199.18/(129-8)=1.65$, which of course should be admitted that there are some uncertainties for the pole position affected by the large errors.
In our formalism, both the states $a_{0}(980)^{+}$ and $a_{0}(1710)^{+}$ are dynamically generated from the coupled channel interactions in pure $S$-wave. 
But, they are also affected by the $P$-wave amplitudes for the states $\bar{K}^{*}(892)^{0}$ and $K^{*}(892)^{+}$, since the values of the phases $\phi_{\bar{K}^{*}(892)^{0}}$ and $\phi_{K^{*}(892)^{+}}$ are in fact $\frac{\pi}{2}\pm 0.1$ (for the central values) as shown in Table \ref{tab:Parameters}, which is just a bit deviated from orthogonality and leads to nonzero interference effect, even though the effect is small. 
Note that, the difference between these two phases are about 0.21, which is close to the experimental measurement but with opposite sign, {\it i.e.}, $-0.16 \pm 0.12 \pm 0.11$ in Ref. \cite{BESIII:2022npc}, within the uncertainties. 
In Fig. \ref{fig:Fig}, our fitting results describe well the data of the three invariant mass distributions \cite{BESIII:2022npc}, where one feature of our fit is only one set parameter used in the combined fitting procedure, as given in Table \ref{tab:Parameters}.
An enhancement at the threshold in the $K_{S}^{0}K^{+}$ mass distribution is caused by the resonance $a_{0}(980)^{+}$ as shown by the dot (magenta) line in Fig. \ref{fig:Fig12}, which is dynamically generated in the $S$-wave final state interactions with the ChUA. The bump structure around $1.25$ GeV in Fig. \ref{fig:Fig12} is the reflection contributions of both the states $\bar{K}^{*}(892)^{0}$ and $K^{*}(892)^{+}$ in the $P$-wave. The obvious peak structure around $1.6$ to $1.8$ GeV in Fig. \ref{fig:Fig12} is contributed by the reflection contributions from the states $\bar{K}^{*}(892)^{0}$ and $K^{*}(892)^{+}$ and the significant signal of the resonance $a_{0}(1710)^{+}$ in the $S$-wave interactions, which comes along with $a_{0}(980)^{+}$ from the coupled channel interaction of one amplitude, see Eq. (\ref{eq:amplitudes2}). 
In Fig. \ref{fig:Fig13} for the $K_{S}^{0}\pi^{0}$ invariant mass distribution, the peak of $\bar{K}^{*}(892)^{0}$ is obvious in the middle-energy region, contrarily, the $S$-wave and $K^{*}(892)^{+}$ contributions are concentrated in the low and high-energy regions. 
Similarly, for the $K^{+}\pi^{0}$ mass distribution in Fig. \ref{fig:Fig23}, except for the peak of the $K^{*}(892)^{+}$, the states $a_{0}(980)^{+}$, $a_{0}(1710)^{+}$, and $\bar{K}^{*}(892)^{0}$ enhance in the energy regions near the threshold, and the state $\bar{K}^{*}(892)^{0}$ also contributes to the enhancement in the high-energy region. Note that as one can see in the low-energy region of Figs. \ref{fig:Fig13} and \ref{fig:Fig23}, there are still some differences between our fit and the experimental data, which, as implied by the experiment, may be caused by the contribution of the resonance $\bar{K}^{*}(1410)^{0}$, not considered in our formalism.

\begin{table}[htbp]
	\centering
	\caption{Values of the parameters from the fit.}
	\resizebox{1.0\textwidth}{!}
	{\begin{tabular}{ccccc}
			\hline \hline 
			Par. &\quad $\mu$&\quad $\mathcal C_{1}$&\quad $\mathcal C_{2}$&\quad $\mathcal C_{3}$   \\
			\hline 
			Fit \MakeUppercase{\romannumeral 1} &\quad $0.716\pm0.013$ GeV&\quad $47518.79\pm7523.18$&\quad $1595.34\pm138.51$&\quad $46454.25\pm3868.04$ \\	
			\hline 	
			&\quad $\mathcal D_{1}$&\quad $\mathcal D_{2}$&\quad $\phi_{\bar{K}^{*}(892)^{0}}$&\quad $\phi_{K^{*}(892)^{+}}$ \\
			\hline 
			&\quad $61.65\pm2.33$&\quad $40.43\pm2.95$&\quad $1.46\pm0.12$&\quad $1.67\pm0.15$ \\
			\hline\hline  
	\end{tabular}}
	\label{tab:Parameters}
\end{table}

\begin{figure}[htbp]
	\begin{subfigure}{0.45\textwidth}
		\centering
		\includegraphics[width=1\linewidth]{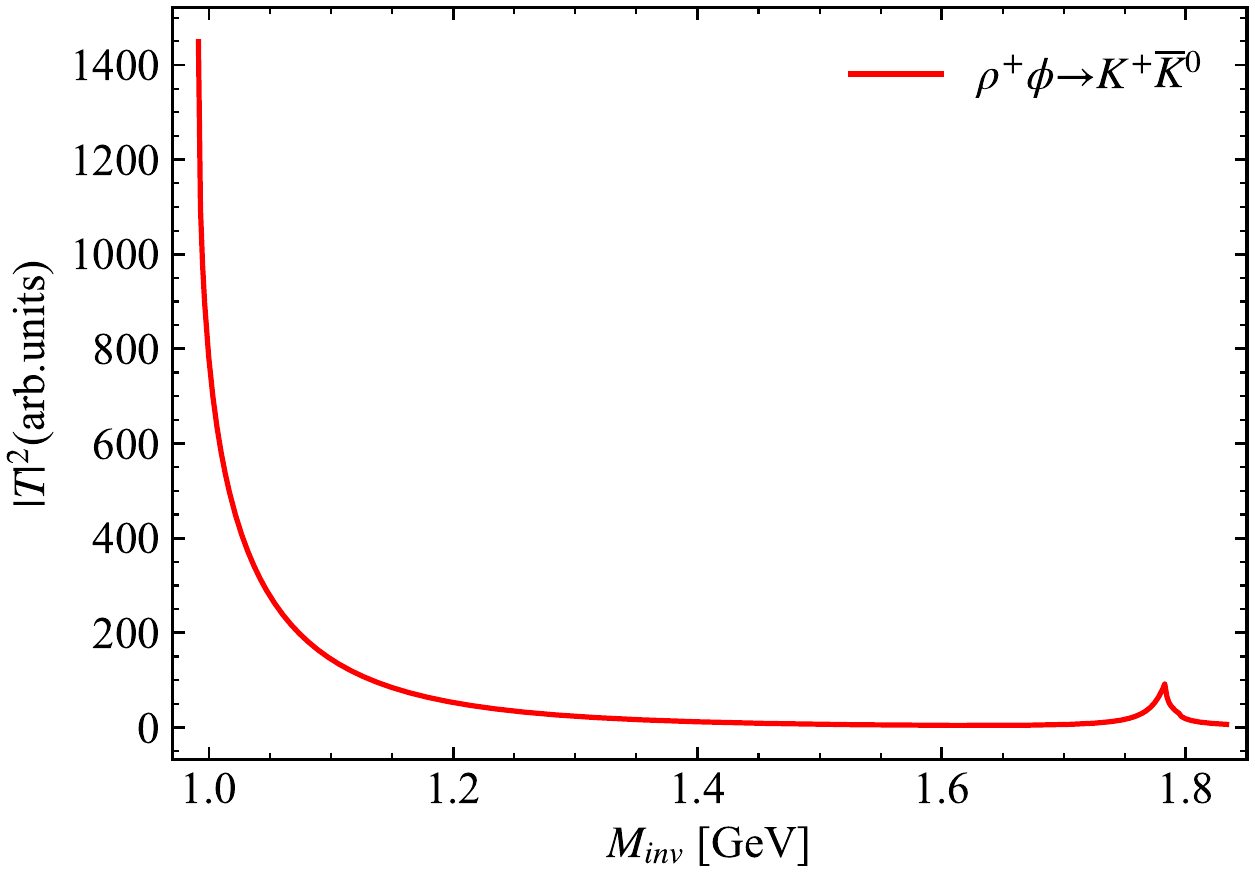} 
		\caption{\footnotesize $\rho^{+}\phi\rightarrow K^{+}\bar{K}^{0}$.}
		\label{fig:T34}
	\end{subfigure}
	\quad
	\quad
	\begin{subfigure}{0.45\textwidth}  
		\centering 
		\includegraphics[width=1\linewidth]{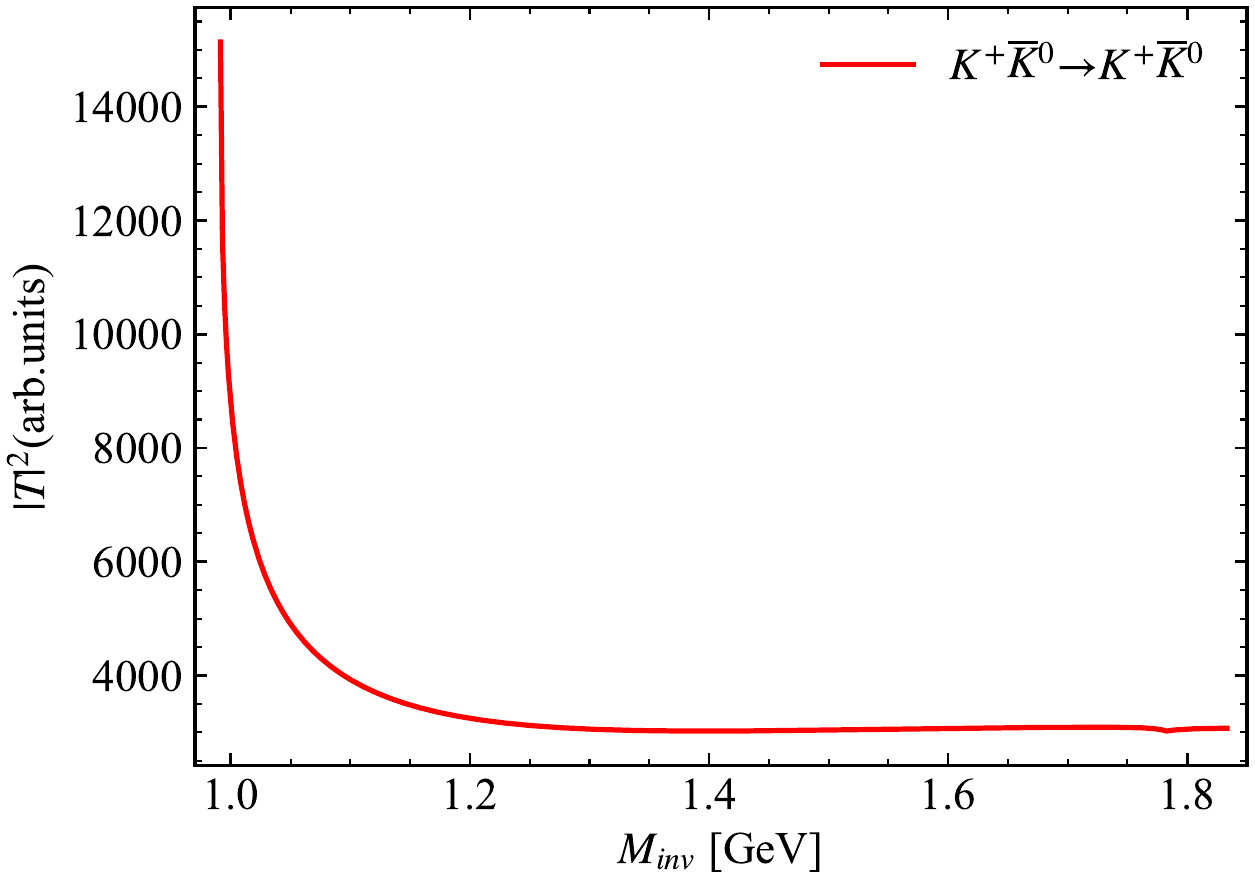} 
		\caption{\footnotesize $K^{+}\bar{K}^{0}\rightarrow K^{+}\bar{K}^{0}$.}
		\label{fig:T44}  
	\end{subfigure}	
	\begin{subfigure}{1\textwidth}  
		\centering 
		\includegraphics[width=0.45\linewidth]{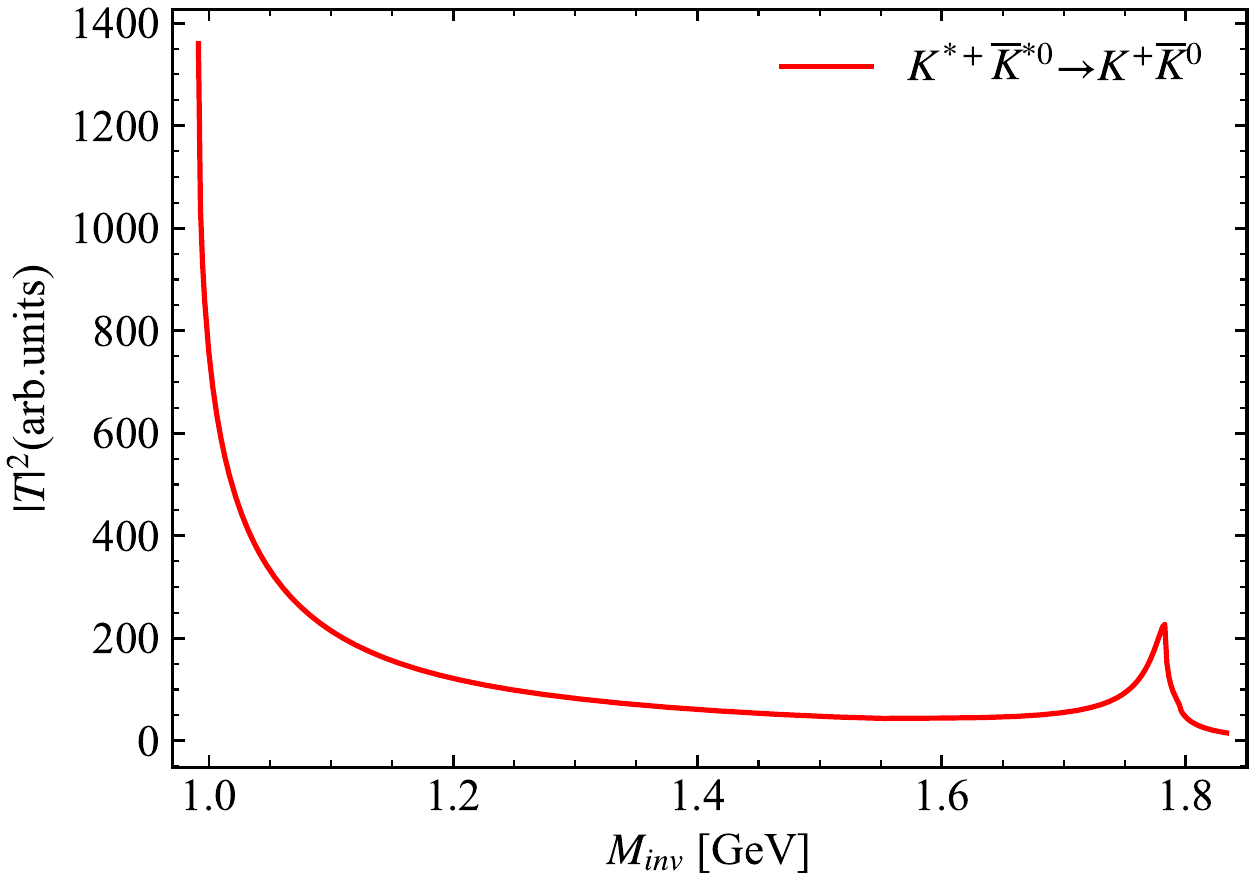} 
		\caption{\footnotesize $K^{*+}\bar{K}^{*0}\rightarrow K^{+}\bar{K}^{0}$.}
		\label{fig:T04}  
	\end{subfigure}	
	\caption{Modulus square of the amplitudes (a) $\rho^{+}\phi\rightarrow K^{+}\bar{K}^{0}$, (b) $K^{+}\bar{K}^{0}\rightarrow K^{+}\bar{K}^{0}$, and (c) $K^{*+}\bar{K}^{*0}\rightarrow K^{+}\bar{K}^{0}$.}
	\label{fig:T4404}
\end{figure}

In Fig. \ref{fig:T4404}, we show the modulus square of the two-body $\rho^{+}\phi\rightarrow K^{+}\bar{K}^{0}$, $K^{+}\bar{K}^{0}\rightarrow K^{+}\bar{K}^{0}$, and $K^{*+}\bar{K}^{*0}\rightarrow K^{+}\bar{K}^{0}$ amplitudes, where the $a_{0}(980)^{+}$ signal near the threshold is strengthened, which is also found in the $K^{+}K^{-}$ invariant mass spectrum of the $D_{s}^{+}\rightarrow K^{+}K^{-}\pi^{+}$ decay including the intermediate resonances $f_{0}(980)$ and $a_{0}(980)$ in the experimental results of Ref. \cite{BESIII:2020ctr}.
One thing should be mentioned that as we discuss in the formalism, the regularization scale $\mu$ is a free parameter in our formalism, which is determined from the fit, and the subtraction constants $a_{ii}(\mu)$ for each channels are evaluated by Eq. (\ref{eq:aii}), different from what had been done in Refs. \cite{Ahmed:2020kmp,Geng:2008gx,Du:2018gyn,Wang:2022pin}.
With the fitting results, the obtained $a_{ii}(\mu)$ have been given in Eq. \eqref{eq:amu}, and the corresponding poles for the states $a_{0}(980)$ and $a_{0}(1710)$ in the complex second Riemann sheets are shown in Table \ref{tab:Poles}. In Table \ref{tab:Poles}, the pole for the state $a_{0}(980)$ is not much different with the ones obatined in Ref. \cite{Ahmed:2020kmp}, which indicates that the interactions of vector meson channels have little influence on this state. 
For the $a_{0}(1710)$, it is obvious that the obtained width is at least seven times smaller than the ones of Refs. \cite{Geng:2008gx,Wang:2022pin} and three times smaller than the one gotten in Ref. \cite{Du:2018gyn}. 
Note that we only evaluate the interactions of the isospin $I=1$ sector for the final state interactions of the decay $D_{s}^{+} \rightarrow K_{S}^{0}K^{+}\pi^{0}$, where one can expect that the states $f_{0}(980)$ and $f_{0}(1710)$ could be reproduced together with the similar two-body interaction formalism in the isospin $I=0$ sector, showing the molecular nature for them.

\begin{table}[htbp]
	\centering
	\caption{Poles \footnote{Note that the poles are always a pair of conjugated solutions in the complex Riemann sheet.} compared with the other works (Unit: GeV).}
	\resizebox{1\textwidth}{!}
	{\begin{tabular}{cccccc}
			\hline \hline 
			&\quad This work &\quad Ref. \cite{Ahmed:2020kmp}&\quad Ref. \cite{Geng:2008gx}&\quad Ref. \cite{Wang:2022pin}&\quad Ref. \cite{Du:2018gyn} \\
			\hline 
			 Par. &\quad $\mu=0.716$&\quad $q_{max}=0.931$, $q_{max}=1.08$&\quad $\mu=1.00$&\quad $q_{max}=1.00$&\quad $q_{max}=1.00$ , $g_{1}=4.596$\\
			\hline 
			$a_{0}(980)$ &\quad $1.0419+0.0345i$&\quad $1.0029+0.0567i$, $0.9745+0.0573i$&\quad $-$&\quad $-$&\quad $-$ \\	
			\hline 
			$a_{0}(1710)$ &\quad $1.7936+0.0094i$&\quad $-$&\quad $1.780-0.066i$&\quad $1.72-0.10i$&\quad $1.76\pm0.03i$ \\	
			\hline\hline  
	\end{tabular}}
	\label{tab:Poles}
\end{table}

Furthermore, based on the results in Table \ref{tab:Poles}, we also concern the widths and partial decay widths of the poles for the corresponding resonances. Since the pole is in fact located at $(M_R + i \frac{\Gamma_R^{tot}}{2})$, one can easily obtain the (total) widths $\Gamma_R^{tot}$ of the corresponding poles for the states $a_{0}(980)$ and $a_{0}(1710)$ from the results in Table \ref{tab:Poles}. For the partial decay widths of each coupled channel, we take the formulae from Refs.  \cite{Oller:1997ti, Oller:1998hw}, written
\begin{equation}
	\begin{aligned}
		\Gamma_{R\to i}=-\frac{1}{16 \pi^2} \int_{E_{\min }}^{E_{\max }} d E \frac{q_{cmi}}{E^2} 4 M_R \operatorname{Im} T_{ii},
	\end{aligned}
	\label{eq:PW3}
\end{equation}
\begin{equation}
	\begin{aligned}
		\Gamma_{R\to j}=-\frac{1}{16 \pi^2} \int_{E_{\min }}^{E_{\max }} d E \frac{q_{cmj}}{E^2} 4 M_R \frac{\left(\operatorname{Im} T_{ji}\right)^2}{\operatorname{Im} T_{ii}},
	\end{aligned}
	\label{eq:PW4}
\end{equation}
where $E$ stands for the total energy of the meson-meson system in the c.m. frame, $q_{cmi}$ ($q_{cmj}$) is the three momentum of the meson in the c.m. frame, given by Eq. \eqref{eq:qcm}, and the amplitudes $T_{ij}$ are evaluated by Eq. \eqref{eq:BSE}. The obtained results are shown in Tables \ref{tab:PDW1_1} and \ref{tab:PDW2_1}. Note that, the $\Gamma_{a_{0}(980)^{+}\to K^{+}\bar{K}^{0}}$ is calculated with Eq. \eqref{eq:PW3}, and the others with Eq. \eqref{eq:PW4}. Meanwhile, the integration limits are taken from threshold to 1.1 GeV for the results in Table \ref{tab:PDW1_1}, and taken from 1.7 to 2.0 GeV for the ones in Table \ref{tab:PDW2_1}. The results in Table \ref{tab:PDW2_1} are somehow very small for $a_{0}(1710)^{+}$ decaying into the channels $K^{+}\bar{K}^{0}$ and $\pi^{+}\eta$, which are also different from the ones predicted in Refs. \cite{Wang:2022pin,Oset:2012zza}. Note that, in Ref. \cite{Geng:2008gx} it was found that the width did not increase much when the contributions of the box diagrams were included, and thus, it was concluded that the predicted $a_{0}(1710)$ state had a small branching ratio to two pseudoscalars.
\begin{table}[htbp]
	\centering
	\setlength{\tabcolsep}{22mm}{
		\caption{The partial decay widths of $a_{0}(980)^{+}$.}
		\resizebox{1.0\textwidth}{!}
		{\begin{tabular}{cc}
				\hline \hline 
				$\Gamma_{a_{0}(980)^{+}\to K^{+}\bar{K}^{0}}$&$\Gamma_{a_{0}(980)^{+}\to \pi^{+}\eta}$  \\
				\hline 
				$28.38$ MeV&$43.60$ MeV \\
				\hline\hline  
		\end{tabular}}
		\label{tab:PDW1_1}}
\end{table}
\begin{table}[htbp]
	\centering
	\setlength{\tabcolsep}{12mm}{
		\caption{The partial decay widths of $a_{0}(1710)^{+}$.}
		\resizebox{1.0\textwidth}{!}
		{\begin{tabular}{ccc}
				\hline \hline 
				$\Gamma_{a_{0}(1710)^{+}\to \rho^{+}\omega}$&$\Gamma_{a_{0}(1710)^{+}\to K^{+}\bar{K}^{0}}$&$\Gamma_{a_{0}(1710)^{+}\to \pi^{+}\eta}$  \\
				\hline 
				$19.65$ MeV&$0.54$ MeV&$0.05$ MeV \\
				\hline\hline  
		\end{tabular}}
		\label{tab:PDW2_1}}
\end{table}

In addition, we also calculate the branching ratios of the corresponding decay channels. For the decays $D_{s}^{+} \rightarrow \bar{K}^{*}(892)^{0}K^{+}$, $K^{*}(892)^{+}K_{S}^{0}$, and $a_{0}(980)^{+}\pi^{0}$, we integrate the three corresponding invariant mass spectra from the threshold to $1.2$ GeV. The uncertainties come from the changes of upper limits $1.20\pm0.05$ GeV. For the $D_{s}^{+} \rightarrow a_{0}(1710)^{+}\pi^{0}$ decay, the integration limits are taken from $1.6$ GeV to ($m_{D_{s}^{+}}-m_{\pi^{0}}$), the uncertainties are from the changes of $1.60\pm0.05$ GeV. The results are given as follows
\begin{equation} 
	\begin{aligned}
		\frac{\mathcal{B}(D_{s}^{+} \rightarrow K^{*}(892)^{+}K_{S}^{0}, K^{*}(892)^{+} \rightarrow K^{+}\pi^{0})}{\mathcal{B}(D_{s}^{+} \rightarrow \bar{K}^{*}(892)^{0}K^{+}, \bar{K}^{*}(892)^{0} \rightarrow K_{S}^{0}\pi^{0})} = 0.40 ^{+0.002}_{-0.003},
	\end{aligned}
	\label{eq:Ratios1}
\end{equation}
\begin{equation} 
	\begin{aligned}
		\frac{\mathcal{B}(D_{s}^{+} \rightarrow a_{0}(980)^{+}\pi^{0}, a_{0}(980)^{+} \rightarrow K_{S}^{0}K^{+})}{\mathcal{B}(D_{s}^{+} \rightarrow \bar{K}^{*}(892)^{0}K^{+}, \bar{K}^{*}(892)^{0} \rightarrow K_{S}^{0}\pi^{0})} = 0.53 ^{+0.06}_{-0.08},
	\end{aligned}
	\label{eq:Ratios2}
\end{equation}
\begin{equation} 
	\begin{aligned}
		\frac{\mathcal{B}(D_{s}^{+} \rightarrow a_{0}(1710)^{+}\pi^{0}, a_{0}(1710)^{+} \rightarrow K_{S}^{0}K^{+})}{\mathcal{B}(D_{s}^{+} \rightarrow \bar{K}^{*}(892)^{0}K^{+}, \bar{K}^{*}(892)^{0} \rightarrow K_{S}^{0}\pi^{0})} = 0.41 ^{+0.04}_{-0.05}.
	\end{aligned}
	\label{eq:Ratios3}
\end{equation}
Then we take the branching fraction $\mathcal{B}(D_{s}^{+} \rightarrow \bar{K}^{*}(892)^{0}K^{+}, \bar{K}^{*}(892)^{0} \rightarrow K_{S}^{0}\pi^{0})=(4.77\pm0.38\pm0.32)\times10^{-3}$ 
\footnote{Note that, only the results for the decays $(D_{s}^{+} \rightarrow \bar {K}^{*}(892)^{0}K^{+}, \bar {K}^{*}(892)^{0} \rightarrow K^{-}\pi^{+})$ and $D_{s}^{+} \rightarrow K^{*}(892)^{+}\bar{K}^{0}$ are found in PDG \cite{Workman:2022ynf}. In principle, with these results in PDG one can obtain the ratio of Eq. \eqref{eq:Ratios1} under the isospin symmetry to the strong decay. But, since the branching fraction of the decay $D_{s}^{+} \rightarrow K^{*}(892)^{+}\bar{K}^{0}$ is evaluated with the results from low statistics, we do not take it into account in the present work.}
measured by the BESIII Collaboration \cite{BESIII:2022npc} as the input, and get the branching ratios for the  other channels, written
\begin{equation} 
	\begin{aligned}
		\mathcal{B}(D_{s}^{+} \rightarrow K^{*}(892)^{+}K_{S}^{0}, K^{*}(892)^{+} \rightarrow K^{+}\pi^{0})=(1.91\pm0.20 ^{+0.01}_{-0.01})\times10^{-3}, \\
		\mathcal{B}(D_{s}^{+} \rightarrow a_{0}(980)^{+}\pi^{0}, a_{0}(980)^{+} \rightarrow K_{S}^{0}K^{+})=(2.53\pm0.26^{+0.27}_{-0.38})\times10^{-3}, \\
		\mathcal{B}(D_{s}^{+} \rightarrow a_{0}(1710)^{+}\pi^{0}, a_{0}(1710)^{+} \rightarrow K_{S}^{0}K^{+})=(1.94\pm0.20^{+0.18}_{-0.24})\times10^{-3}, \\
	\end{aligned}
	\label{eq:Theory}
\end{equation}
where the first uncertainties are estimated from the experimental errors, and the second ones come from the Eqs. (\ref{eq:Ratios1}-\ref{eq:Ratios3}). The following results are taken from the experimental measurements \cite{BESIII:2022npc},
\begin{equation} 
	\begin{aligned}
		\mathcal{B}(D_{s}^{+} \rightarrow K^{*}(892)^{+}K_{S}^{0}, K^{*}(892)^{+} \rightarrow K^{+}\pi^{0})=(2.03\pm0.26\pm0.20)\times10^{-3}, \\
		\mathcal{B}(D_{s}^{+} \rightarrow a_{0}(980)^{+}\pi^{0}, a_{0}(980)^{+} \rightarrow K_{S}^{0}K^{+})=(1.12\pm0.25\pm0.27)\times10^{-3}, \\
		\mathcal{B}(D_{s}^{+} \rightarrow a_{0}(1710)^{+}\pi^{0}, a_{0}(1710)^{+} \rightarrow K_{S}^{0}K^{+})=(3.44\pm0.52\pm0.32)\times10^{-3}. \\
	\end{aligned}
	\label{eq:BESIII}
\end{equation}
Compared with the experimental measurements of Eq. \eqref{eq:BESIII}, our results of the branching fractions in Eq. \eqref{eq:Theory} for the decay $D_{s}^{+} \rightarrow K^{*}(892)^{+}K_{S}^{0}$ is a little smaller, but they are consistent with each other within the uncertainties. 
Whereas, the one for the decay $D_{s}^{+} \rightarrow a_{0}(980)^{+}\pi^{0}$ is two times bigger than the measurement result. 
For the decay $D_{s}^{+} \rightarrow a_{0}(1710)^{+}\pi^{0}$, our result is 1/3 smaller than the experimental measurement. 
However, note that, in Ref. \cite{Dai:2021owu} the predicted branching ratio of $D_{s}^{+} \rightarrow a_{0}(1710)^{+}\pi^{0}$ is $(1.3\pm0.4)\times10^{-3}$, which is smaller than what we have.
 
\section{Conclusions}
\label{sec:Conclusions}

We study the weak decay process of $D_{s}^{+} \rightarrow K_{S}^{0}K^{+}\pi^{0}$ by considering the mechanisms of external and internal $W$-emission in the quark level. 
In the hadron level, based on the final state interaction formalism, including the contributions of tree-level and rescattering of the interactions $\rho^{+}\phi \rightarrow K^{+}\bar{K}^{0}$, $K^{+}\bar{K}^{0}\rightarrow K^{+}\bar{K}^{0}$, and $K^{*+}\bar{K}^{*0}\rightarrow K^{+}\bar{K}^{0}$, the $K_{S}^{0}K^{+}$ invariant mass spectrum is described with the main contributions from the resonances $a_{0}(980)^{+}$ and $a_{0}(1710)^{+}$. 
Note that these two states are dynamically reproduced with the chiral unitary approach, where the coupled channel interactions including the pseudoscalar and vector channels are taken into account coherently. 
Moreover, combining with the $P$-wave contributions from the states $\bar{K}^{*}(892)^{0}$ and $K^{*}(892)^{+}$, the experimental data of the three mass distributions in the decay $D_{s}^{+} \rightarrow K_{S}^{0}K^{+}\pi^{0}$ are well described, where it can be found that the reflections of these states are important to the spectra as shown in Fig. \ref{fig:Fig}, and one should keep in mind that only one set of free parameter is used in the combined fit. 
In addition, with the fitted regularization scale $\mu$ for deteriming $a_{ii}(\mu)$ by Eq. (\ref{eq:aii}), we find the poles of the states $a_{0}(980)^{+}$ and $a_{0}(1710)^{+}$ in the corresponding Riemann sheets, which are consistent with the results in Refs. \cite{Ahmed:2020kmp,Geng:2008gx,Wang:2022pin,Du:2018gyn}, except for a bit small width of the $a_{0}(1710)^{+}$. 
Our results indicate that the $a_{0}(980)$ is a $K\bar K$ bound state, and the $a_{0}(1710)$ is a $K^{*}\bar K^{*}$ bound state. 
Furthermore, we evaluate the branching ratios of related decay channels. Within the uncertainties, the obtained results are consistent with the experimental measurements in the magnitudes. 
In view of these results, the state found in the $K_{S}^{0}K^{+}$ invariant mass spectrum is indeed the $a_{0}(1710)$, not a new $a_{0}(1817)$ state. \\

Note added: When our manuscript is prepared, one work on the decay $D_{s}^{+} \rightarrow K_{S}^{0}K^{+}\pi^{0}$ is given in Ref. \cite{Zhu:2022guw}, of which the formalism is similar. But, in the present work, both the resonances $a_{0}(980)^{+}$ and $a_{0}(1710)^{+}$ are dynamically generated in the coupled channel interactions.

\section*{Acknowledgements}

We would like to thank Xiang Liu, Zhi-Feng Sun, Si-Qiang Luo, Zheng-Li Wang, and Yu Lu for valuable discussions. 
This work is supported by the National Natural Science Foundation of China (NSFC) under Grant No. 12247101 (ZYW), 
and partly by the Natural Science Foundation of Changsha under Grant No. kq2208257 and the Natural Science Foundation of Hunan province under Grant No. 2023JJ30647 (CWX).

 \addcontentsline{toc}{section}{References}

\begin{thebibliography}{9}

%
\bibitem{BESIII:2022npc}
M.~Ablikim \textit{et al.} [BESIII],
Phys. Rev. Lett. \textbf{129}, no.18, 18 (2022)
[arXiv:2204.09614 [hep-ex]].

\bibitem{CLEO:2013bae}
P.~U.~E.~Onyisi \textit{et al.} [CLEO],
Phys. Rev. D \textbf{88}, no.3, 032009 (2013)
[arXiv:1306.5363 [hep-ex]].

\bibitem{BaBar:2021fkz}
J.~P.~Lees \textit{et al.} [BaBar],
Phys. Rev. D \textbf{104}, no.7, 072002 (2021)
[arXiv:2106.05157 [hep-ex]].

\bibitem{BaBar:2018uqa}
J.~P.~Lees \textit{et al.} [BaBar],
Phys. Rev. D \textbf{97}, no.11, 112006 (2018)
[arXiv:1804.04044 [hep-ex]].

\bibitem{BESIII:2021anf}
M.~Ablikim \textit{et al.} [BESIII],
Phys. Rev. D \textbf{105}, no.5, L051103 (2022)
[arXiv:2110.07650 [hep-ex]].

\bibitem{Godfrey:1985xj}
S.~Godfrey and N.~Isgur,
Phys. Rev. D \textbf{32}, 189-231 (1985)

\bibitem{Segovia:2008zza}
J.~Segovia, D.~R.~Entem and F.~Fernandez,
Phys. Lett. B \textbf{662}, 33-36 (2008)

\bibitem{Chanowitz:2005du}
M.~Chanowitz,
Phys. Rev. Lett. \textbf{95}, 172001 (2005)
[arXiv:hep-ph/0506125 [hep-ph]].

\bibitem{Chao:2007sk}
K.~T.~Chao, X.~G.~He and J.~P.~Ma,
Phys. Rev. Lett. \textbf{98}, 149103 (2007)
[arXiv:0704.1061 [hep-ph]].

\bibitem{Close:2005vf}
F.~E.~Close and Q.~Zhao,
Phys. Rev. D \textbf{71}, 094022 (2005)
[arXiv:hep-ph/0504043 [hep-ph]].

\bibitem{Giacosa:2005zt}
F.~Giacosa, T.~Gutsche, V.~E.~Lyubovitskij and A.~Faessler,
Phys. Rev. D \textbf{72}, 094006 (2005)
[arXiv:hep-ph/0509247 [hep-ph]].

\bibitem{Cheng:2006hu}
H.~Y.~Cheng, C.~K.~Chua and K.~F.~Liu,
Phys. Rev. D \textbf{74}, 094005 (2006)
[arXiv:hep-ph/0607206 [hep-ph]].

\bibitem{Albaladejo:2008qa}
M.~Albaladejo and J.~A.~Oller,
Phys. Rev. Lett. \textbf{101}, 252002 (2008)
[arXiv:0801.4929 [hep-ph]].

\bibitem{Gui:2012gx}
L.~C.~Gui \textit{et al.} [CLQCD],
Phys. Rev. Lett. \textbf{110}, no.2, 021601 (2013)
[arXiv:1206.0125 [hep-lat]].

\bibitem{Janowski:2014ppa}
S.~Janowski, F.~Giacosa and D.~H.~Rischke,
Phys. Rev. D \textbf{90}, no.11, 114005 (2014)
[arXiv:1408.4921 [hep-ph]].

\bibitem{Ochs:2013gi}
W.~Ochs,
J. Phys. G \textbf{40}, 043001 (2013)
[arXiv:1301.5183 [hep-ph]].

\bibitem{Cheng:2015iaa}
H.~Y.~Cheng, C.~K.~Chua and K.~F.~Liu,
Phys. Rev. D \textbf{92}, no.9, 094006 (2015)
[arXiv:1503.06827 [hep-ph]].

\bibitem{Klempt:2021wpg}
E.~Klempt and A.~V.~Sarantsev,
Phys. Lett. B \textbf{826}, 136906 (2022)
[arXiv:2112.04348 [hep-ph]].

\bibitem{BESIII:2022riz}
M.~Ablikim \textit{et al.} [BESIII],
Phys. Rev. Lett. \textbf{129}, no.19, 192002 (2022)
[arXiv:2202.00621 [hep-ex]].

\bibitem{BESIII:2022iwi}
M.~Ablikim \textit{et al.} [BESIII],
Phys. Rev. D \textbf{106}, no.7, 072012 (2022)
[arXiv:2202.00623 [hep-ex]].

\bibitem{Oller:1997ti}
J.~A.~Oller and E.~Oset,
Nucl. Phys. A \textbf{620}, 438-456 (1997)
[erratum: Nucl. Phys. A \textbf{652}, 407-409 (1999)]
[arXiv:hep-ph/9702314 [hep-ph]].

\bibitem{Oset:1997it}
E.~Oset and A.~Ramos,
Nucl. Phys. A \textbf{635}, 99-120 (1998)
[arXiv:nucl-th/9711022 [nucl-th]].

\bibitem{Oller:2000ma}
J.~A.~Oller, E.~Oset and A.~Ramos,
Prog. Part. Nucl. Phys. \textbf{45}, 157-242 (2000)
[arXiv:hep-ph/0002193 [hep-ph]].

\bibitem{Oller:2000fj}
J.~A.~Oller and U.-G.~Mei{\ss}ner,
Phys. Lett. B \textbf{500}, 263-272 (2001)
[arXiv:hep-ph/0011146 [hep-ph]].

\bibitem{Oset:2008qh}
E.~Oset, L.~S.~Geng, D.~Gamermann, M.~J.~Vicente Vacas, D.~Strottman, K.~P.~Khemchandani, A.~Martinez Torres, J.~A.~Oller, L.~Roca and M.~Napsuciale,
Int. J. Mod. Phys. E \textbf{18}, 1389-1403 (2009)
[arXiv:0806.0340 [nucl-th]].

\bibitem{Geng:2008gx}
L.~S.~Geng and E.~Oset,
Phys. Rev. D \textbf{79}, 074009 (2009)
[arXiv:0812.1199 [hep-ph]].

\bibitem{Du:2018gyn}
M.~L.~Du, D.~G\"ulmez, F.~K.~Guo, U.-G.~Mei{\ss}ner and Q.~Wang,
Eur. Phys. J. C \textbf{78}, no.12, 988 (2018)
[arXiv:1808.09664 [hep-ph]].

\bibitem{Wang:2021jub}
Z.~L.~Wang and B.~S.~Zou,
Phys. Rev. D \textbf{104}, no.11, 114001 (2021)
[arXiv:2107.14470 [hep-ph]].

\bibitem{Wang:2022pin}
Z.~L.~Wang and B.~S.~Zou,
Eur. Phys. J. C \textbf{82}, no.6, 509 (2022)
[arXiv:2203.02899 [hep-ph]].

\bibitem{Guo:2017jvc}
F.~K.~Guo, C.~Hanhart, U.-G.~Mei{\ss}ner, Q.~Wang, Q.~Zhao and B.~S.~Zou,
Rev. Mod. Phys. \textbf{90}, no.1, 015004 (2018)
[erratum: Rev. Mod. Phys. \textbf{94}, no.2, 029901 (2022)]
[arXiv:1705.00141 [hep-ph]].

\bibitem{BESIII:2020ctr}
M.~Ablikim \textit{et al.} [BESIII],
Phys. Rev. D \textbf{104}, no.1, 012016 (2021)
[arXiv:2011.08041 [hep-ex]].

\bibitem{Dai:2021owu}
L.~R.~Dai, E.~Oset and L.~S.~Geng,
Eur. Phys. J. C \textbf{82}, no.3, 225 (2022)
[arXiv:2111.10230 [hep-ph]].

\bibitem{Zhu:2022wzk}
X.~Zhu, D.~M.~Li, E.~Wang, L.~S.~Geng and J.~J.~Xie,
Phys. Rev. D \textbf{105}, no.11, 116010 (2022)
[arXiv:2204.09384 [hep-ph]].

\bibitem{Guo:2022xqu}
D.~Guo, W.~Chen, H.~X.~Chen, X.~Liu and S.~L.~Zhu,
Phys. Rev. D \textbf{105}, no.11, 114014 (2022)
[arXiv:2204.13092 [hep-ph]].

\bibitem{BES:2006vdb}
M.~Ablikim \textit{et al.} [BES],
Phys. Rev. Lett. \textbf{96}, 162002 (2006)
[arXiv:hep-ex/0602031 [hep-ex]].

\bibitem{Chau:1982da}
L.~L.~Chau,
Phys. Rept. \textbf{95}, 1-94 (1983)

\bibitem{Chau:1987tk}
L.~L.~Chau and H.~Y.~Cheng,
Phys. Rev. D \textbf{36}, 137 (1987)

\bibitem{Liang:2015qva}
W.~H.~Liang, J.~J.~Xie and E.~Oset,
Eur. Phys. J. C \textbf{75}, no.12, 609 (2015)
[arXiv:1510.03175 [hep-ph]].

\bibitem{Ahmed:2020qkv}
H.~A.~Ahmed, Z.~Y.~Wang, Z.~F.~Sun and C.~W.~Xiao,
Eur. Phys. J. C \textbf{81}, no.8, 695 (2021)
[arXiv:2011.08758 [hep-ph]].

\bibitem{Liang:2014tia}
W.~H.~Liang and E.~Oset,
Phys. Lett. B \textbf{737}, 70-74 (2014)
[arXiv:1406.7228 [hep-ph]].

\bibitem{Molina:2019udw}
R.~Molina, J.~J.~Xie, W.~H.~Liang, L.~S.~Geng and E.~Oset,
Phys. Lett. B \textbf{803}, 135279 (2020)
[arXiv:1908.11557 [hep-ph]].

\bibitem{BESIII:2019jjr}
M.~Ablikim \textit{et al.} [BESIII],
Phys. Rev. Lett. \textbf{123}, no.11, 112001 (2019)
[arXiv:1903.04118 [hep-ex]].

\bibitem{Hsiao:2019ait}
Y.~K.~Hsiao, Y.~Yu and B.~C.~Ke,
Eur. Phys. J. C \textbf{80}, no.9, 895 (2020)
[arXiv:1909.07327 [hep-ph]].

\bibitem{Ling:2021qzl}
X.~Z.~Ling, M.~Z.~Liu, J.~X.~Lu, L.~S.~Geng and J.~J.~Xie,
Phys. Rev. D \textbf{103}, no.11, 116016 (2021)
[arXiv:2102.05349 [hep-ph]].


\bibitem{Wang:2019niy}
Z.~L.~Wang and B.~S.~Zou,
Phys. Rev. D \textbf{99}, no.9, 096014 (2019)
[arXiv:1901.10169 [hep-ph]].

\bibitem{Duan:2020vye}
M.~Y.~Duan, J.~Y.~Wang, G.~Y.~Wang, E.~Wang and D.~M.~Li,
Eur. Phys. J. C \textbf{80}, no.11, 1041 (2020)
[arXiv:2008.10139 [hep-ph]].

\bibitem{Xie:2014tma}
J.~J.~Xie, L.~R.~Dai and E.~Oset,
Phys. Lett. B \textbf{742}, 363-369 (2015)
[arXiv:1409.0401 [hep-ph]].

\bibitem{Bando:1984ej}
M.~Bando, T.~Kugo, S.~Uehara, K.~Yamawaki and T.~Yanagida,
Phys. Rev. Lett. \textbf{54}, 1215 (1985)

\bibitem{Bando:1987br}
M.~Bando, T.~Kugo and K.~Yamawaki,
Phys. Rept. \textbf{164}, 217-314 (1988)

\bibitem{Molina:2008jw}
R.~Molina, D.~Nicmorus and E.~Oset,
Phys. Rev. D \textbf{78}, 114018 (2008)
[arXiv:0809.2233 [hep-ph]].

\bibitem{Oset:2012zza}
E.~Oset, L.~S.~Geng and R.~Molina,
J. Phys. Conf. Ser. \textbf{348}, 012004 (2012)

\bibitem{Oller:1998zr}
J.~A.~Oller and E.~Oset,
Phys. Rev. D \textbf{60}, 074023 (1999)
[arXiv:hep-ph/9809337 [hep-ph]].

\bibitem{Gamermann:2006nm}
D.~Gamermann, E.~Oset, D.~Strottman and M.~J.~Vicente Vacas,
Phys. Rev. D \textbf{76}, 074016 (2007)
[arXiv:hep-ph/0612179 [hep-ph]].

\bibitem{Alvarez-Ruso:2010rqm}
L.~Alvarez-Ruso, J.~A.~Oller and J.~M.~Alarcon,
Phys. Rev. D \textbf{82}, 094028 (2010)
[arXiv:1007.4512 [hep-ph]].

\bibitem{Guo:2016zep}
Z.~H.~Guo, L.~Liu, U.-G.~Mei{\ss}ner, J.~A.~Oller and A.~Rusetsky,
Phys. Rev. D \textbf{95}, no.5, 054004 (2017)
[arXiv:1609.08096 [hep-ph]].

\bibitem{Wang:2021kka}
Z.~Y.~Wang, H.~A.~Ahmed and C.~W.~Xiao,
Phys. Rev. D \textbf{105}, no.1, 016030 (2022)
[arXiv:2110.05359 [hep-ph]].

\bibitem{Guo:2018tjx}
Z.~H.~Guo, L.~Liu, U.~G.~Mei\ss{}ner, J.~A.~Oller and A.~Rusetsky,
Eur. Phys. J. C \textbf{79}, no.1, 13 (2019)
doi:10.1140/epjc/s10052-018-6518-1
[arXiv:1811.05585 [hep-ph]].

\bibitem{Toledo:2020zxj}
G.~Toledo, N.~Ikeno and E.~Oset,
Eur. Phys. J. C \textbf{81}, no.3, 268 (2021)
[arXiv:2008.11312 [hep-ph]].

\bibitem{Roca:2020lyi}
L.~Roca and E.~Oset,
Phys. Rev. D \textbf{103}, no.3, 034020 (2021)
[arXiv:2011.05185 [hep-ph]].

\bibitem{Workman:2022ynf}
R.~L.~Workman [Particle Data Group],
PTEP \textbf{2022}, 083C01 (2022)

\bibitem{Wang:2021ews}
Z.~Y.~Wang, J.~Y.~Yi, Z.~F.~Sun and C.~W.~Xiao,
Phys. Rev. D \textbf{105}, no.1, 016025 (2022)
[arXiv:2109.00153 [hep-ph]].

\bibitem{Ahmed:2020kmp}
H.~A.~Ahmed and C.~W.~Xiao,
Phys. Rev. D \textbf{101}, no.9, 094034 (2020)
[arXiv:2001.08141 [hep-ph]].

\bibitem{Oller:1998hw}
J.~A.~Oller, E.~Oset and J.~R.~Pelaez,
Phys. Rev. D \textbf{59}, 074001 (1999)
[erratum: Phys. Rev. D \textbf{60}, 099906 (1999); erratum: Phys. Rev. D \textbf{75}, 099903 (2007)]
[arXiv:hep-ph/9804209 [hep-ph]].

\bibitem{Zhu:2022guw}
X.~Zhu, H.~N.~Wang, D.~M.~Li, E.~Wang, L.~S.~Geng and J.~J.~Xie,
Phys. Rev. D \textbf{107}, no.3, 034001 (2023)
[arXiv:2210.12992 [hep-ph]].

\end{thebibliography}
\end{document}